\documentclass[12pt]{iopart}

\usepackage{iopams}  
\usepackage{color}
\usepackage[dvipsnames]{xcolor} 
\usepackage{tikz}
\usetikzlibrary{decorations.pathmorphing,patterns} 
\usepackage{psfrag}
\usepackage{graphicx}
\usepackage{stackrel}
\usepackage{amsfonts}
\usepackage{bm}          
\usepackage{amssymb}
\usepackage{perpage}
\usepackage{siunitx}
\MakePerPage{footnote}

\newcommand{\be}{\begin{equation}}
\newcommand{\ee}{\end{equation}}
\newcommand{\ba}{\begin{eqnarray}}
\newcommand{\ea}{\end{eqnarray}}

\begin{document}

\title[Critical percolation on the kagome hypergraph]
{Critical percolation on the kagome hypergraph}

\author{Christian R.\ Scullard$^{1}$, Jesper Lykke Jacobsen$^{2,3,4}$, and Robert M. Ziff$^{5}$}
\address{${}^1$Lawrence Livermore National Laboratory, Livermore CA 94550, USA}
\address{$^2$Laboratoire de Physique de l'\'Ecole Normale Sup\'erieure, ENS, Universit\'e PSL, CNRS, Sorbonne Universit\'e, Universit\'e de Paris, Paris, France}
\address{$^3$Sorbonne Universit\'e, \'Ecole Normale Sup\'erieure, CNRS, Laboratoire de Physique (LPENS), 75005 Paris, France} 
\address{$^4$Institut de Physique Th\'eorique, Universit\'e Paris Saclay, CEA, CNRS, 91191 Gif-sur-Yvette, France}
\address{${}^5$Center for the Study of Complex Systems and Department of Chemical Engineering, Ann Arbor, Michigan 48109-2136, USA}

\eads{\mailto{scullard1@llnl.gov}, \mailto{jesper.jacobsen@ens.fr}, \mailto{rziff@umich.edu}}

\begin{abstract}
We study the percolation critical surface of the kagome lattice in which each triangle is allowed an arbitrary connectivity. Using the method of critical polynomials, we find points along this critical surface to high precision. This kagome hypergraph contains many unsolved problems as special cases, including bond percolation on the kagome and $(3,12^2)$ lattices, and site percolation on the hexagonal, or honeycomb, lattice, as well as a single point for which there is an exact solution. We are able to compute enough points along the critical surface to find a very accurate fit, essentially a Taylor series about the exact point, that allows estimations of the critical point of any system that lies on the surface to precision rivaling Monte Carlo and traditional techniques of similar accuracy. We find also that this system sheds light on some of the surprising aspects of the method of critical polynomials, such as why it is so accurate for certain problems, like the kagome and $(3,12^2)$ lattices. The bond percolation critical points of these lattices can be found to 17 and 18 digits, respectively, because they are in close proximity, in a sense that can be made quantitative, to the exact point on the critical surface. We also discuss in detail a parallel implementation of the method which we use here for a few calculations.
\end{abstract}

\noindent

\section{Introduction}
Percolation \cite{Grimmett} is a fundamental model in the statistical mechanics of phase transitions. The original lattice model has spawned many generalizations, such as continuum \cite{Meester}, explosive \cite{Ziff2010,Achlioptas}, and gradient \cite{Rosso} variants. In the standard bond percolation model, the edges of a given lattice in dimension $d$ are independently declared open with probability $p$ and closed with probability $1-p$. For an infinite lattice, there is a critical probability, $p_c$, above which there is an infinite cluster and below which there is none. In dimension greater than one, the problem of locating $p_c$ exactly is unsolved in general. In two dimensions there is a narrow class of lattices for which the problem can be solved \cite{SykesEssam,Wierman84,Scullard06,Ziff06} but for finite $d>2$, no critical points are known exactly.

\begin{figure}
\begin{center}
\includegraphics{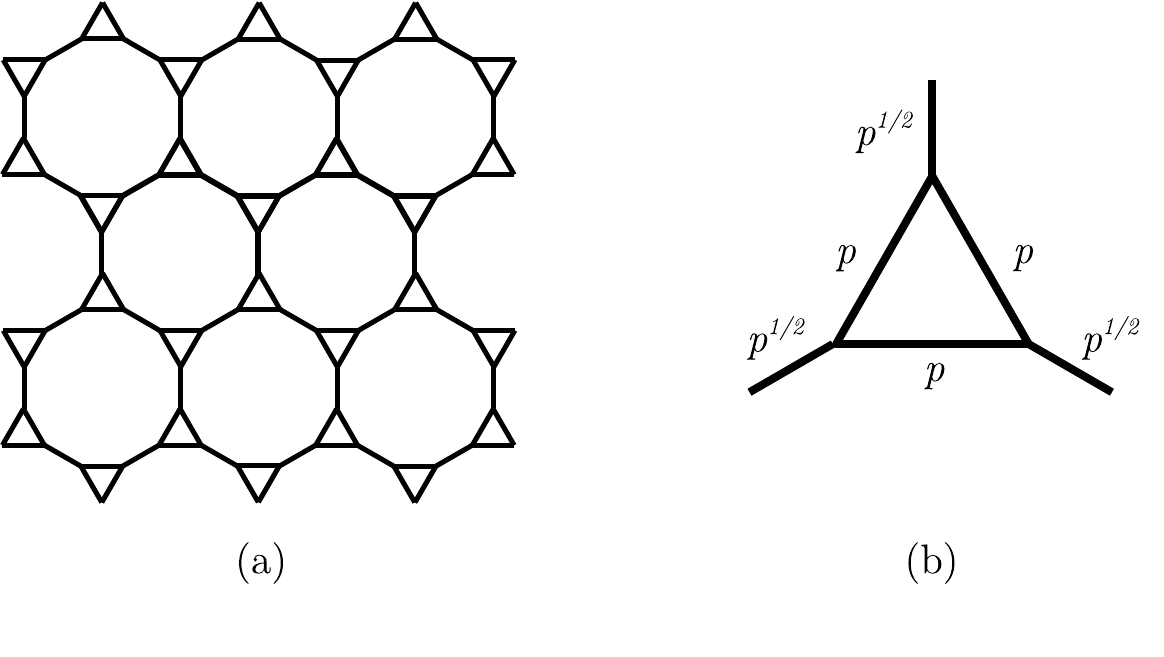}
\caption{(a) The $(3,12^2)$ lattice; (b) the triangular cell for this problem. }
\label{fig:3-12}
\end{center}
\end{figure}

Here, we discuss a version of two-dimensional percolation on the kagome lattice (figure \ref{fig:hypergraphs}b) in which each of its triangles, rather than containing three simple bonds, supports a general arrangement in which $P_0$, $P_2$, and $P_3$ denote respectively the probability that none of its three vertices is connected inside the triangle, that only two of its vertices are connected, and that all three are connected. There is then a critical surface for this problem characterized by the function $P_3(P_0)$, which gives the critical value of $P_3$ corresponding to some arbitrarily chosen $P_0$, $P_2$ having been eliminated by normalization. This critical surface has previously been explored by one of us \cite{ZiffGu09} using Monte Carlo techniques. We revisit the problem here with the far more accurate method of critical polynomials. This method has recently been used to find the bond percolation thresholds of all the unsolved Archimedean lattices \cite{ScullardJacobsen2020} to accuracy far outstripping what is practically possible with Monte Carlo. For example, in the case of the $(3,12^2)$ lattice, figure \ref{fig:3-12}a, it was found that
\begin{equation}
 p_c(3,12^2)= 0.740\,420\,798\,850\,811\,610(2) \label{eq:312pc}
\end{equation}
compared with the Monte Carlo result $p_c= 0.74042077(2)$ \cite{Ding10}. One of the goals of the present work is to explain how this seeming fanciful accuracy actually arises. And in fact, we have several other reasons for studying this system. First, it is a kind of template, containing many unsolved problems as special cases, such as bond percolation on the kagome and $(3,12^2)$ lattices and site percolation on the hexagonal lattice. Secondly, we can use this system to study the finite-size scaling of critical polynomial estimates, a subject for which we currently have little theoretical understanding. And third, as mentioned, it allows us to explain why exactly the critical polynomial method provides such excellent estimates for the kagome and $(3,12^2)$ lattices, namely their proximity to the one exact solution on the critical surface of the kagome hypergraph.

The paper is organized as follows. First, we discuss hypergraph systems beginning with triangular hypergraphs, for which there is an exact solution for the whole critical surface, and describe the kagome hypergraph in detail. We then review the method of critical polynomials and explain how we apply it to the hypergraph problem and outline the construction of the transfer matrix necessary for the calculations. 

\section{Exact critical surfaces}
A triangular hypergraph system in shown in figure \ref{fig:hypergraphs}a. Within each shaded triangle we might have any network of bonds or sites, including correlated bonds and internal sites, or we may have a system that is not easily represented as a graph. A percolation process may be defined on this hypergraph by defining the quantities $P_3$, the probability that all three vertices within a triangle are connected, $P_2$, the probability that exactly two are connected and $P_0$, the probability that none are connected. Further demanding the triangle be isotropic (i.e., the three potentially different $P_2$ are equal), normalization requires
\begin{equation}
 P_0+3 P_2+P_3=1 ,
\end{equation}
and thus we are free to choose only $P_0$ and $P_3$. For every choice of $P_0$ in an acceptable range, there is a value of $P_3$ at which the system is critical. It is well known that for the triangular hypergraph system,  this critical surface is given by \cite{Scullard06,Ziff06,Wu06}
\begin{equation}
 P_3=P_0 , \label{eq:critcond}
\end{equation}
provided $P_3$ and $P_0$ are identical on every triangle. Note that this surface is only defined for $P_0 \in [0,1/2]$ as $P_2$ would be negative outside this range.

Many well-known systems can be identified along this critical surface. For example, the endpoint $P_0=P_3=1/2$ is critical site percolation on the triangular lattice. If the system consists of bond percolation on individual edges, we have
\begin{eqnarray}
 P_0&=&(1-p)^3 \label{eq:triP0} \\
 P_2&=&p(1-p)^2 \\
 P_3&=&p^3+3p^2(1-p) \label{eq:triP3} .
\end{eqnarray}
Applying (\ref{eq:critcond}) gives $p_c=2 \sin \pi/18$ \cite{SykesEssam}, and thus critical triangular bond percolation corresponds to $P_0=0.278\,066\,14...$ .

There are some other examples of hypergraph systems for which (\ref{eq:critcond}) gives the critical point \cite{ZiffScullard06}, but for most\footnote{The crucial point for exact solvability is the self-duality (in the sense of a hypergraph) of the lattice.}, even lattices consisting entirely of triangles (so-called three-regular hypergraphs), it does not. One is then naturally led to wonder what $P_3(P_0)$ might look like in other cases, particularly for systems containing unsolved problems.

\section{Kagome hypergraph}
\begin{figure}
\begin{center}
\includegraphics{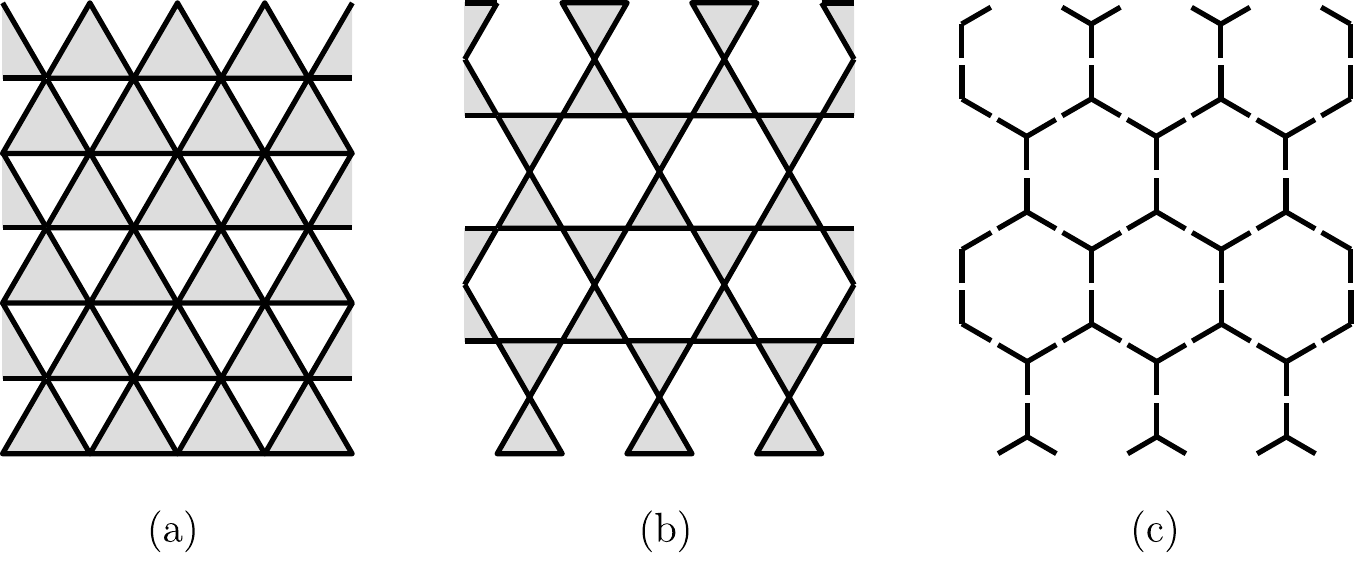}
\caption{a) triangular hypergraph; b) kagome hypergraph; c) the double-bond hexagonal lattice, a special case of b). The shaded triangles can have arbitrary $P_0$ and $P_3$, which are not necessarily realizable in bonds and sites.}
\label{fig:hypergraphs}
\end{center}
\end{figure}

The system we study here is the kagome hypergraph, shown in figure \ref{fig:hypergraphs}b, which has as special cases a wide range of problems for which the critical points are unknown exactly, such as bond percolation on the kagome and $(3,12^2)$ lattices and site percolation on the hexagonal lattice.

However, our goal here is to study the critical manifold of the kagome hypergraph as a function of the probability $P_0$, and the usual control parameter, $p$, on a particular realization of shaded triangles will therefore take a somewhat secondary role. Specifically, it is the case that for every $P_0$ in a suitable range, there is a critical $P_3$ above which there is an infinite cluster and below which there is none. Like the problem of ordinary bond percolation on the kagome lattice, the determination of $P_3(P_0)$ is an unsolved problem for general $P_0$. There is however one special case in which the exact solution is known. If the shaded triangle is a star of three bonds, the result is the double-bond hexagonal (DBH) lattice, shown in figure \ref{fig:hypergraphs}(c) for which the critical probability is known \cite{SykesEssam} to be
\begin{equation}
 p_c^{\mathrm{DBH}}=\sqrt{1-2 \sin \frac{\pi}{18}} = 0.807\,900\,76...\ ,
\end{equation}
but for our purposes we refer to this critical point by its value of $P_0$, which we denote $P_0^*$,
\begin{equation}
 P_0^*=(1-p_c^{\mathrm{DBH}})^3+3(1-p_c^{\mathrm{DBH}})^2 p_c^{\mathrm{DBH}} = 0.096\,528\,61... \ .
\end{equation}
At $P_0^*$, the critical value of $P_3$, $P_3^*$, can likewise be calculated,
\begin{equation}
 P_3^*=(p_c^{\mathrm{DBH}})^3 = 0.527\,319\,77... \ .
\end{equation}
The rest of the range of $P_3(P_0)$ contains only unsolved problems whose critical points are known only numerically. However, as already mentioned, the method of critical polynomials has located these probabilities for two of these problems, the kagome and $(3,12^2)$ lattices, to 17 and 18 digits, respectively. The goal of the present work is to compute the rest of the curve, not necessarily to that accuracy, but certainly to greater accuracy than is currently possible with Monte Carlo.

\section{Critical polynomials}
The method of critical polynomials \cite{Scullard11,Scullard11-2,Jacobsen12,ScullardJacobsen2012,Jacobsen14b,ScullardJacobsen16} has now been used extensively on a variety of percolation problems, and, as pointed out in the introduction, was recently used \cite{ScullardJacobsen2020} in massively parallel computations to obtain the bond percolation critical points of the 8 unsolved Archimedean lattices. The idea is simple and we discuss here only how the method will apply to the kagome hypergraph. First, one chooses a basis for the lattice, $B$, consisting of a finite subgraph with periodic boundary conditions. Then on $B$ one computes $\mathrm{P}_{\mathrm{2D}}(P_0,P_3;B)$ and $\mathrm{P}_{\mathrm{0D}}(P_0,P_3;B)$, the probability that there is an infinite cluster that spans all copies of the basis and the probability that no cluster spans either direction, respectively. We then set \cite{ScullardJacobsen2012}
\begin{equation}
 \mathrm{P}_{\mathrm{2D}}(P_0,P_3;B)=\mathrm{P}_{\mathrm{0D}}(P_0,P_3;B) \label{eq:2D0D}
\end{equation}
which gives an implicit relationship between $P_3$ and $P_0$. For any finite $B$ this will not be exact in general, although it will be exact for systems of the type in figure \ref{fig:hypergraphs}(a) because in that case (\ref{eq:2D0D}) reduces to (\ref{eq:critcond}). When (\ref{eq:2D0D}) is not exact, it can be improved by taking progressively larger $B$.

\begin{figure}
\begin{center}
\includegraphics[width=1.25in]{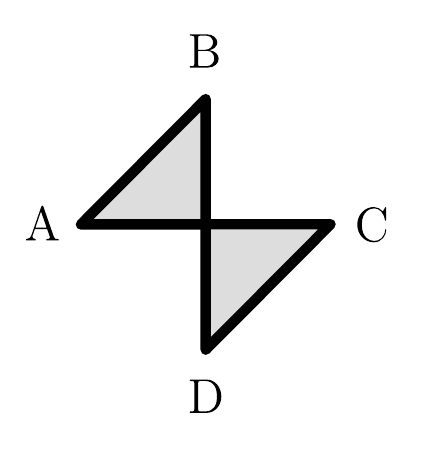}
\caption{Embedding of a unit cell of the kagome hypergraph into the squares of figure \ref{fig:square-basis-loop}. This, along with the definitions in figure \ref{fig:squares} lead to the weights in \ref{sec:weights}. An alternative choice would be to flip this figure horizontally (or vertically with the same result), which would lead to a different set of weights.}
\label{fig:kagome_square}
\end{center}
\end{figure}

Let us take for a basis a single unit cell of the kagome hypergraph, as shown in figure \ref{fig:kagome_square}. One then easily finds
\begin{eqnarray}
 \mathrm{P}_{\mathrm{2D}}(P_0,P_3;B) &=& P_3^2\\ 
 \mathrm{P}_{\mathrm{0D}}(P_0,P_3;B) &=& 2 P_3 P_0+P_0^2+6 P_0 P_2+6 P_2^2
\end{eqnarray}
and, using the normalization condition to eliminate $P_2$ and solving for $P_3$, we find
\begin{equation}
 P_3(P_0)=-2+2 P_0+\sqrt{3}\sqrt{2-2P_0+P_0^2} , \label{eq:firstpoly}
\end{equation}
which is valid for $P_0 \in [0,(2-\sqrt{2})/2]$. This is not expected to be exact anywhere except at the point $P_0^*$, the critical point of the DBH, where we do indeed recover the correct $P_3^*$. Over the rest of the range of (\ref{eq:firstpoly}), the predicted value of $P_3$ is an approximation. For example, if we use the simple expressions for three bonds in a triangle with probability $p$ in (\ref{eq:triP0}) and (\ref{eq:triP3}) for $P_0$ and $P_3$, then we have the ordinary kagome lattice and plugging these into (\ref{eq:firstpoly}) we find $p_c$ is the solution to the equation
\begin{equation}
 1 - 3 p^2 - 6 p^3 + 12 p^4 - 6 p^5 + p^6 = 0
\end{equation}
with the solution
\begin{equation}
 p_c=0.524\,429\,7... , \label{eq:kagomepoly}.
\end{equation}
This is only an approximation to the true value \cite{ScullardJacobsen2020}, 
\begin{equation}
p_c=0.524\,404\,999\,167\,448\,20(1) \label{eq:kagome_num}
\end{equation}
albeit accurate to four figures. Note that the approximation (\ref{eq:kagomepoly}) is identical to the one found by Wu in 1979 \cite{Wu79} using his ``homogeneity'' conjecture.

As another example, place inside each triangle a smaller triangle of three bonds with weight p, as well as three bonds with weight $p^{1/2}$ connecting the smaller triangle to the corners as in figure \ref{fig:3-12}b. The result is the $(3,12^2)$ lattice, where the $p^{1/2}$ compensate for the fact that some of its bonds are doubled (in series). After a bit of drawing one finds:
\begin{eqnarray}
P_3 &=& p^{9/2} + 3 p^{7/2} (1-p) \\
P_2 &=& p^2 (1-p)^2 (1-\sqrt{p}) + 3 p^3(1-p)(1-\sqrt{p}) \cr
 & & + p^{5/2}(1-p)^2 + p^4(1-\sqrt{p}) \\
P_0 &=& 1 - P_3 - 3 P_2
\end{eqnarray}
Plugging this into (\ref{eq:firstpoly}) and solving produces
\begin{equation}
p_c = 0.740\,423\,3 \cdots \label{eq:312poly}
\end{equation}
which agrees with the value (\ref{eq:312pc}) to 5 digits.

A third example is the site hexagonal problem, in which each triangle is completely connected with probability $p$ and disconnected with probability $1-p$, i.e.,
\begin{eqnarray}
 P_3 &=&p \\
 P_2 &=&0 \\
 P_0 &=&1-p .
\end{eqnarray}
In this case, plugging into (\ref{eq:firstpoly}) yields $p_c=1/\sqrt{2}=0.707\,107...$, whereas traditional numerical methods for this system give \cite{FengDengBlote08}
\begin{equation}
 p_c=0.697\,040\,24(4), \label{eq:hex_site_num}
\end{equation}
so the first polynomial is in reasonable agreement, but to nowhere near the accuracy of the former two examples.

\begin{figure}
\begin{center}
\includegraphics[width=5in]{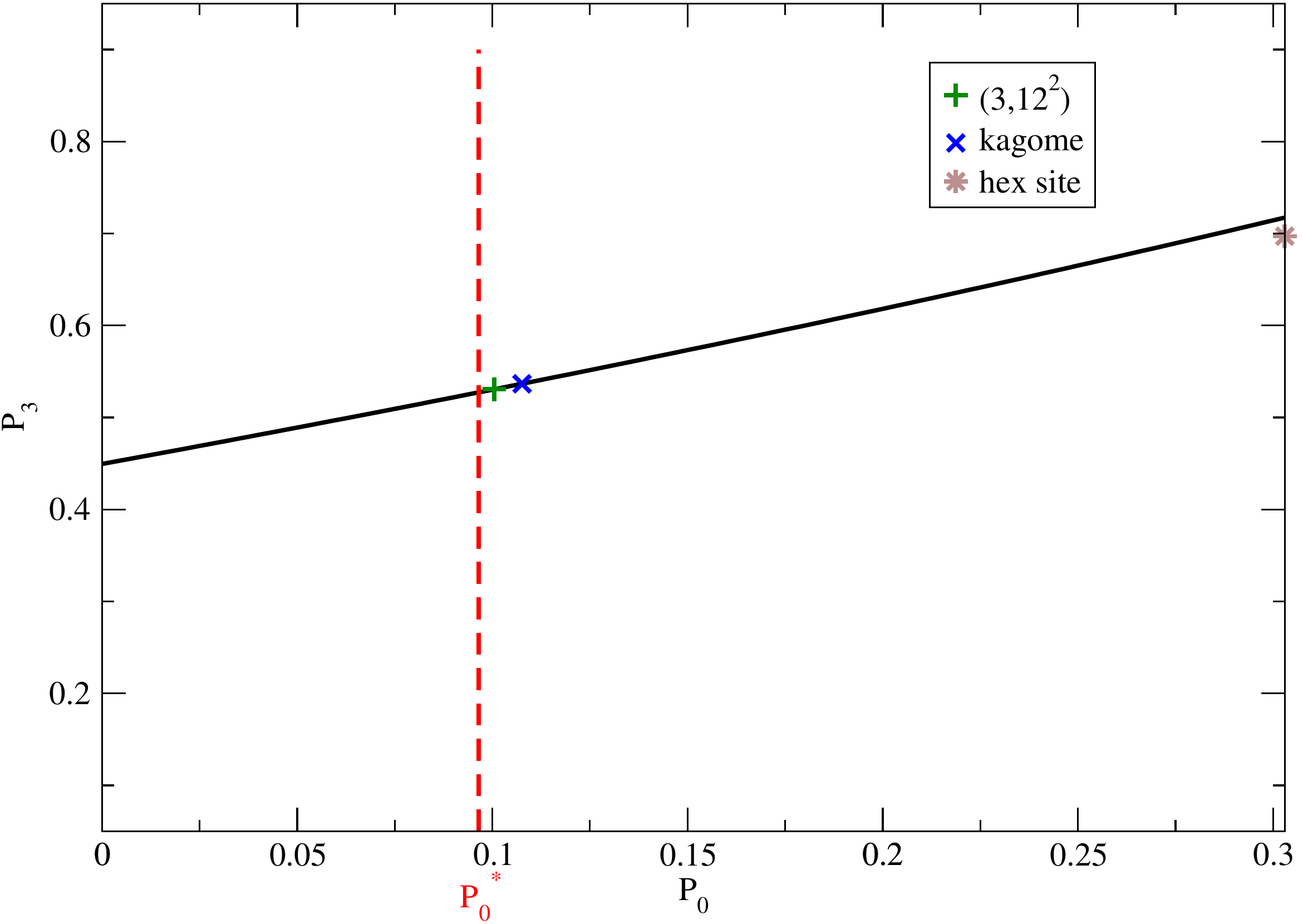}
\caption{The critical curve $P_3(P_0)$ calculated with the first polynomial approximation (\ref{eq:firstpoly}), along with more precise data points for the $(3,12^2)$, kagome and site hexagonal problems. Near the exact point, $P_0^*$, the approximation is indistinguishable on the plot from the more accurate calculations. It is less accurate at the extreme end of the range but is still good to within 2\%.}
\label{fig:P3}
\end{center}
\end{figure}

The curve (\ref{eq:firstpoly}) is plotted in figure \ref{fig:P3} along with the numerically calculated points $(P_0,P_3)$ of the critical kagome, $(3,12^2)$ and site hexagonal problems. One of the conclusions of this work will be that the accuracy of the polynomial approximation depends strongly on a system's proximity to the exact solution, $P_0^*$. The increase of precision for the $(3,12^2)$ lattice in eq.\ (\ref{eq:312poly}) over eq.\ (\ref{eq:kagomepoly}) for the kagome lattice (i.e., improving from 4 to 5 correct digits) illustrates well the location of the coloured points in figure \ref{fig:P3}, i.e. $(3,12^2)$ is closer to exact solvability. By the same token, the distance of the critical site hexagonal problem value of $P_0$ from $P_0^*$ results in a larger deviation from the first polynomial estimate. This point will be made more clearly in section \ref{sec:derivatives}.

Note that all systems composed of independent sites and bonds appear to have $P_0>P_0^*$. Systems with $P_0<P_0^*$ seemingly necessarily include correlations of some kind. For example, the system $P_0=0$ is one in which each triangle can only have either two or all three vertices connected, which is not realizable with independent bonds and sites.

\subsection{Transfer matrix}
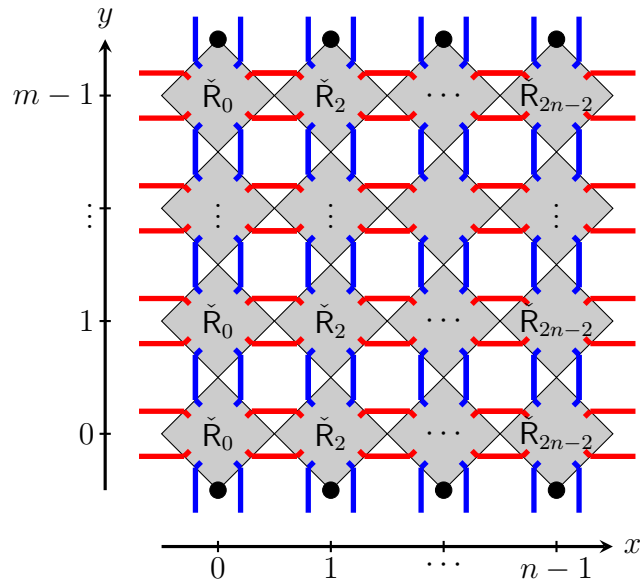
\begin{figure}
\begin{center}

\begin{tikzpicture}[scale=1.5,>=stealth]
\foreach \xpos in {0,1,2,3}
\foreach \ypos in {0,1,2,3}
 \fill[black!20] (\xpos+0.5,\ypos) -- (\xpos+1,\ypos+0.5) -- (\xpos+0.5,\ypos+1) -- (\xpos,\ypos+0.5) -- cycle;
\foreach \xpos in {0,1,2,3}
\foreach \ypos in {0,1,2,3}
 \draw[black] (\xpos+0.5,\ypos) -- (\xpos+1,\ypos+0.5) -- (\xpos+0.5,\ypos+1) -- (\xpos,\ypos+0.5) -- cycle;

\foreach \xpos in {0,1,2,3}
{
 \draw[fill] (\xpos+0.5,0) circle(0.4ex);
 \draw[fill] (\xpos+0.5,4) circle(0.4ex);
}


\foreach \xpos in {0,1,2,3}
\foreach \ypos in {0,1,2,3}
{
 \draw[red,line width=2pt] (\xpos-0.2,\ypos+0.3) -- (\xpos+0.2,\ypos+0.3);
 \draw[red,line width=2pt] (\xpos+0.8,\ypos+0.3) -- (\xpos+1.2,\ypos+0.3);
 \draw[red,line width=2pt] (\xpos-0.2,\ypos+0.7) -- (\xpos+0.2,\ypos+0.7);
 \draw[red,line width=2pt] (\xpos+0.8,\ypos+0.7) -- (\xpos+1.2,\ypos+0.7);
}

\foreach \xpos in {0,1,2,3}
\foreach \ypos in {0,1,2,3}
{
 \draw[blue,line width=2pt] (\xpos+0.3,\ypos-0.2) -- (\xpos+0.3,\ypos+0.2);
 \draw[blue,line width=2pt] (\xpos+0.7,\ypos-0.2) -- (\xpos+0.7,\ypos+0.2);
 \draw[blue,line width=2pt] (\xpos+0.3,\ypos+0.8) -- (\xpos+0.3,\ypos+1.2);
 \draw[blue,line width=2pt] (\xpos+0.7,\ypos+0.8) -- (\xpos+0.7,\ypos+1.2);
}

\foreach \xpos in {0,1,2,3}
\foreach \ypos in {0,1,2,3}
{
 \draw[red,line width=2pt] (\xpos+0.2,\ypos+0.3) -- (\xpos+0.25,\ypos+0.35);
 \draw[red,line width=2pt] (\xpos+0.8,\ypos+0.3) -- (\xpos+0.75,\ypos+0.35);
 \draw[red,line width=2pt] (\xpos+0.2,\ypos+0.7) -- (\xpos+0.25,\ypos+0.65);
 \draw[red,line width=2pt] (\xpos+0.8,\ypos+0.7) -- (\xpos+0.75,\ypos+0.65);
 \draw[blue,line width=2pt] (\xpos+0.3,\ypos+0.2) -- (\xpos+0.35,\ypos+0.25);
 \draw[blue,line width=2pt] (\xpos+0.7,\ypos+0.2) -- (\xpos+0.65,\ypos+0.25);
 \draw[blue,line width=2pt] (\xpos+0.3,\ypos+0.8) -- (\xpos+0.35,\ypos+0.75);
 \draw[blue,line width=2pt] (\xpos+0.7,\ypos+0.8) -- (\xpos+0.65,\ypos+0.75);
}

\foreach \ypos in {0,1,3}
  \draw (0.5,\ypos+0.5) node{$\check{\sf R}_0$};
\foreach \ypos in {0,1,3}
  \draw (1.5,\ypos+0.5) node{$\check{\sf R}_2$};
\foreach \ypos in {0,1,3}
  \draw (2.5,\ypos+0.5) node{$\cdots$};
\foreach \ypos in {0,1,3}
  \draw (3.5,\ypos+0.5) node{$\check{\sf R}_{2n-2}$};
\foreach \xpos in {0,1,3}
  \draw (\xpos+0.5,2.5) node{$\vdots$};

\draw[very thick,->] (0,-0.5)--(4,-0.5);
\draw (4,-0.5) node[right] {$x$};
\foreach \xpos in {0,1,2,3}
 \draw[thick] (\xpos+0.5,-0.55)--(\xpos+0.5,-0.45);
\draw (0.5,-0.5) node[below] {$0$};
\draw (1.5,-0.5) node[below] {$1$};
\draw (2.5,-0.5) node[below] {$\cdots$};
\draw (3.5,-0.5) node[below] {$n-1$};

\draw[very thick,->] (-0.5,0)--(-0.5,4);
\draw (-0.5,4) node[above] {$y$};
\foreach \ypos in {0,1,2,3}
 \draw[thick] (-0.55,\ypos+0.5)--(-0.45,\ypos+0.5);
\draw (-0.5,0.5) node[left] {$0$};
\draw (-0.5,1.5) node[left] {$1$};
\draw (-0.5,2.5) node[left] {$\vdots$};
\draw (-0.5,3.5) node[left] {$m-1$};
 
\end{tikzpicture}
\caption{Basis of size $n \times m$ in the loop representation. The loop ends are shown in blue (called the ``quantum space'' in \cite{Jacobsen15}) and the auxiliary space is in red. The matrix $\check{\sf R}_i$ adds each grey square.}
 \label{fig:square-basis-loop}
\end{center}
\end{figure}

The central problem of the critical polynomial method is constructing the quantities $P_{\mathrm{2D}}$ and $P_{\mathrm{0D}}$ for bases as large as possible. This is done most efficiently with the transfer matrix. In earlier work \cite{ScullardJacobsen2012,ScullardJacobsen16,JacobsenScullard2013}, the basis was taken to be $n \times m$, as shown in figure \ref{fig:square-basis-loop}; i.e., they were finite in both dimensions. In \cite{Jacobsen15} it was shown by one of us how to take the $m \rightarrow \infty$ limit by recasting the problem as one of finding the largest eigenvalues in two different sectors of the transfer matrix, which is the approach we take here. We operate on the same type of basis as in figure \ref{fig:square-basis-loop}, a square hypergraph in which each shaded square can represent any network of sites and bonds. The kagome hypergraph is a special case of this system, with the shaded squares replaced by the cell in figure \ref{fig:kagome_square}. 

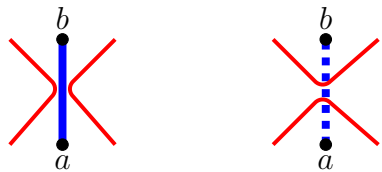
\begin{figure}
\begin{center}
\begin{tikzpicture}[scale=0.7]
 \draw[blue,line width=3pt] (1,-1)--(1,1);
 \draw[fill] (1,-1) circle(0.6ex) node[below] {$a$};
 \draw[fill] (1,1) circle(0.6ex) node[above] {$b$};
 \draw[red,line width=1.5pt] (0,1)--(0.8,0.2) arc(45:-45:2mm) -- (0,-1);
 \draw[red,line width=1.5pt] (2,1)--(1.2,0.2) arc(135:225:2mm) -- (2,-1);

\begin{scope}[xshift=5cm]
 \draw[blue,line width=3pt,dashed] (1,-1)--(1,1);
 \draw[fill] (1,-1) circle(0.6ex) node[below] {$a$};
 \draw[fill] (1,1) circle(0.6ex) node[above] {$b$};
 \draw[red,line width=1.5pt] (0,1)--(0.8,0.2) arc(225:315:2mm) -- (2,1);
 \draw[red,line width=1.5pt] (0,-1)--(0.8,-0.2) arc(135:45:2mm) -- (2,-1);
\end{scope}
\end{tikzpicture}
\caption{Correspondence between a cluster configuration and a loop model. On the left, the edge connecting vertices $a$ and $b$ is present, and the loop segments are drawn on either side of the edge. On the right, the edge is absent and the two loop segments run between the vertices.}
\label{fig:looptrans}
\end{center}
\end{figure}

\begin{figure}
\begin{center}
\includegraphics{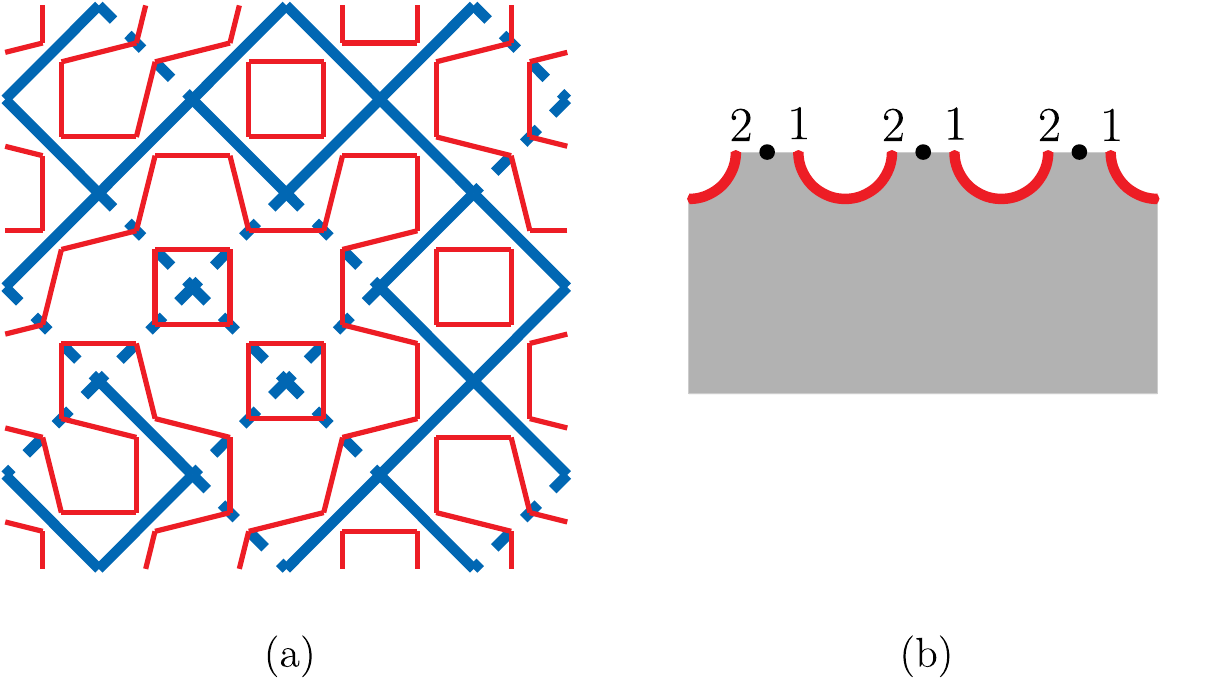}
\caption{(a) The correspondence between a cluster model on a $3 \times 3$ square lattice (heavy blue lines) and a loop model on another square lattice (the corresponding medial lattice; thin red lines). The lattice is periodic in the horizontal direction and we eventually take the vertical direction periodic in the limit of large $m$. (b) Schematic representation of the loop end configuration on the top row of (a), which is the basic data needed for the transfer matrix computation. }
\label{fig:clusters_loops}
\end{center}
\end{figure}

Our transfer matrix operates in the loop representation of the $q$-state Potts model \cite{BaxterKellandWu1976}, of which percolation is the $q \rightarrow 1$ limit. The details were laid out in a paper by one of us \cite{Jacobsen15}, and, although we review them briefly now, the reader is referred there for a fuller explanation. The Potts partition function with edge weight $v$ on a lattice $\mathcal{L}$ is given by \cite{FK1972}
\begin{equation}
 Z = \sum_{A \subseteq E} v^{|A|}q^{k(A)}
\end{equation}
where the sum is over all subsets, $A$, of the set of edges, $E$, $k(A)$ is the number of connected components including isolated vertices, and $|A|$ is the number of edges present in $A$. An example configuration on the square lattice is shown in heavy blue lines in figure \ref{fig:clusters_loops}a. Rather than this cluster representation, we can transform to a loop representation \cite{BaxterKellandWu1976} by employing the rules illustrated in figure \ref{fig:looptrans}, namely drawing different loop segments around an edge depending on whether it is present in or absent from a configuration. The result is a configuration of loops on the medial lattice, $\mathcal{L}_M$, which for the square lattice is another square lattice. The partition function is now given by
\begin{equation}
 Z=q^{|V|/2} \sum_{A \subseteq E}x^{|A|} n_{\mathrm{loop}}^{\ell(A)}
\end{equation}
where $n_{\mathrm{loop}}=\sqrt{q}$ is called the loop fugacity, $x \equiv v/\sqrt{q}$, $|V|$ is the number of vertices, and $\ell(A)$ is the number of closed loops. Note that we could alternatively take the sum to be over polygonal decompositions of $\mathcal{L}_M$ rather than edge subsets on $\mathcal{L}$. In figure \ref{fig:clusters_loops}a, we show in thin red lines the loop configuration corresponding to the cluster configuration in heavy blue.

The basic information needed for a transfer matrix calculation on a periodic lattice strip, or cylinder, of width $n$, consists of the Potts weights (alternatively, the probabilities) of every state of connectivity between the vertices on the top row (we take the transfer direction to be upward). These are represented as components of a vector, ${\bf v}$, and one finds that upon adding a row of the lattice, the new vector of weights, ${\bf v}'$, is the result of a matrix multiplication ${\bf v}'={\mathrm T} {\bf v}$. Here, we are interested in two topological sectors of the transfer matrix: one that contains the 2D wrapping cluster, and one that has no wrapping clusters, which we refer to as 0D. Each sector then has its own set of possible states. For example, in the 2D sector when $n=2$, there are three possible states; one where both vertices are in the 2D cluster, one where only the left, and another where only the right, vertex is in the cluster. As the calculation proceeds, we must make sure that there is always at least one vertex in the 2D cluster. In the 0D sector, there are also three states; one in which the two vertices are disconnected, and two in which they are in the same cluster but connected in different ways (either through the periodic direction or not). In this sector, we must make sure we do not create a 1D wrapping cluster, which would occur if we were to join two vertices of the same cluster that are already connected through the periodic direction. We build up each row one square at a time. Therefore, while the new row in partially completed, there are two extra vertices that must be inserted and which are then removed when the row is finished. These extra vertices are termed {\it auxiliary spaces}.

\begin{figure}
\begin{center}
\includegraphics[width=5in]{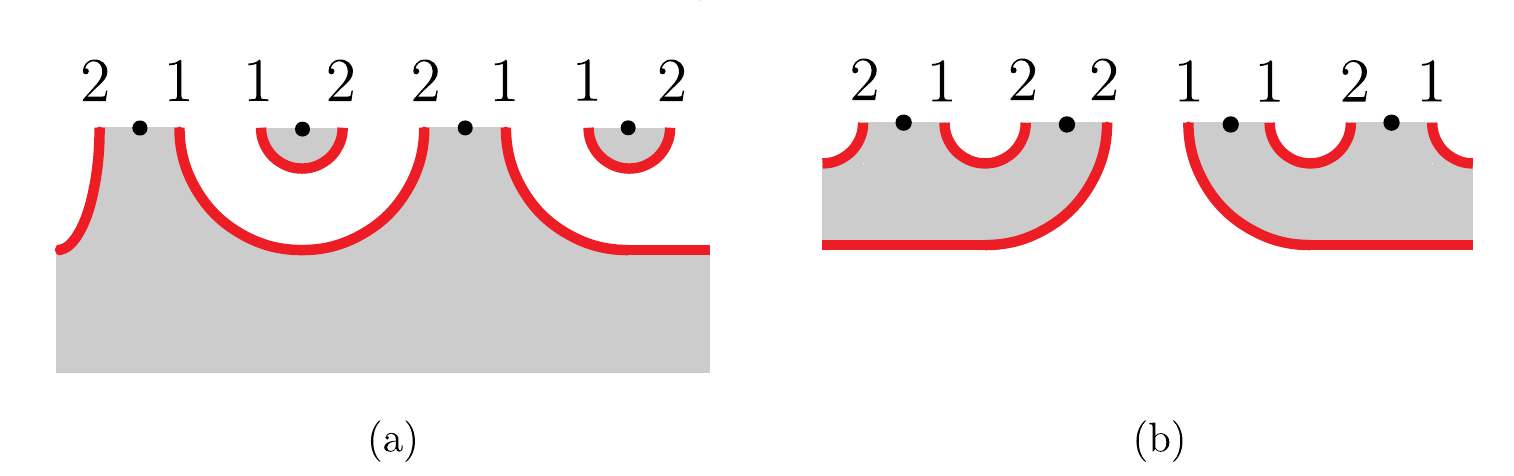}
\caption{(a) An open loop configuration; (b) a closed configuration. }
\label{fig:loops}
\end{center}
\end{figure}

This is the essential idea of the transfer matrix calculation, but in practice we work in the loop representation where each vertex is replaced by two loop ends on either side of it. The basic information is then the connectivity state of the ends of the loops on the top row. In figure \ref{fig:clusters_loops}b, we show the top-row connectivities corresponding to figure \ref{fig:clusters_loops}. There are $2n$ loop ends on the top row, and the state is encoded by assigning half of them the number 1, to indicate the left end of a loop, and 2 to indicate the right. On a planar lattice this assignment to each loop end uniquely specifies the connectivity. See Figure \ref{fig:loops} for further examples.

We have, for a given $n$, two different matrices, $\mathrm{T}_{\mathrm{2D}}$ and $\mathrm{T}_{\mathrm{0D}}$, that each build up their corresponding topological sector on the periodic lattice strip, or cylinder, of width $n$ squares. To approach our problem, we choose a value of $P_0$, and then adjust $P_3$ until the largest eigenvalues of $\mathrm{T}_{\mathrm{2D}}$ and $\mathrm{T}_{\mathrm{0D}}$, $\Lambda_{\mathrm{2D}}$ and $\Lambda_{\mathrm{0D}}$, satisfy \cite{Jacobsen15}
\begin{equation}
 \Lambda_{\mathrm{2D}}(P_0,P_3)=\Lambda_{\mathrm{0D}}(P_0,P_3),
\end{equation}
which determines the values of $P_3$ at which the two probabilities $P_{\mathrm{2D}}$ and $P_{\mathrm{0D}}$ are equal. Eigenvalues are found with power iteration and adjustments are made to $P_3$ using either the Newton-Raphson or Householder method. Isolating the 2D and 0D sectors is simple. To be more precise than above, the transfer matrices $\mathrm{T}_{\mathrm{2D}}$ and $\mathrm{T}_{\mathrm{0D}}$ are pieces of the full transfer matrix, $\mathrm{T}$, in a sense made clear in \cite{Jacobsen15}. Loop configurations can be divided into two types, open and closed. In figure \ref{fig:loops}a, we show an example of an open configuration, in which the region occupied by the cluster, namely the shaded grey area, is unbounded. In figure \ref{fig:loops}b is an example of a closed configuration, where the cluster is confined. The open (closed) configurations correspond to the 2D (0D) sector as long as we do not allow any 1D configurations in which a cluster wraps only the periodic direction. We therefore disallow operations that would give rise to such a situation, such as when joining the fourth and fifth loop ends in figure \ref{fig:loops}b. With this rule in mind, we find the eigenvalue $\Lambda_{\mathrm{2D}}$ ($\Lambda_{\mathrm{0D}}$) by beginning with an open (closed) loop configuration and repeatedly operating with $\mathrm{T}$ until convergence.

Rather than build the lattice one edge at a time, as in Refs.\ \cite{Jacobsen14,Jacobsen15,ScullardJacobsen2020}, here we add an entire square at a time. For a partially completed row, we still need the auxiliary spaces, described above. In the loop representation the two vertices are instead two sets of two loop ends that are inserted into the problem to allow us to add the squares; each square then has four ``incoming'' and four ``outgoing'' loop ends, as shown in Figure \ref{fig:square-basis-loop}. When a row is completed the auxiliary space loop ends are removed by applying periodic boundary conditions (i.e. identifying the loop ends that are on the left and right boundaries).

\begin{figure}
\begin{center}
\includegraphics{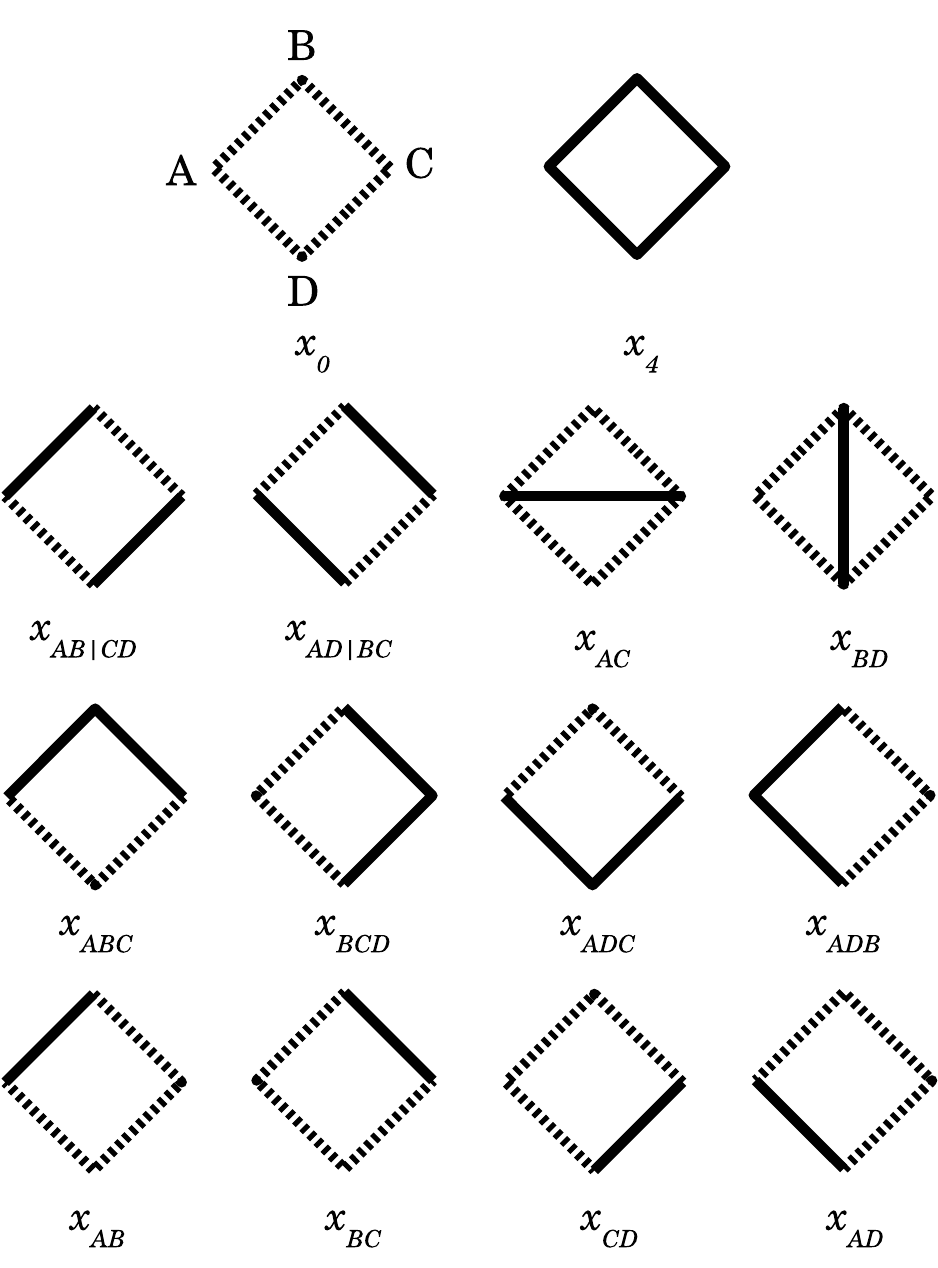}
\caption{The 14 planar connections on a square with their associated weights.}
\label{fig:squares}
\end{center}
\end{figure}

In Refs.\ \cite{Jacobsen14,Jacobsen15}, where the lattice was built up one edge at a time, the addition of a single edge could be written in terms of the Temperley-Lieb (TL) operator, $e_i$ \cite{TemperleyLieb71}. Our operation adding a whole square turns out to be a combination of products of TL operators. We must demand that all connections in a square be planar, otherwise the loop picture cannot be maintained. There are 14 such planar connectivities possible on a square and they are illustrated in Figure \ref{fig:squares} along with the labels for their weights.

This is a very general setup, capable of handling a wide range of problems merely by making different choices for these weights. For example, if we set $x_4=p$ and $x_0=1-p$ and all other $x=0$, we have site percolation on the square lattice with probability $p$. In Figure \ref{fig:kagome_square}, we indicate how the kagome hypergraph is embedded into the square hypergraph and in \ref{sec:weights}, we give the corresponding weights as functions of $P_0$, $P_2$, and $P_3$.

\begin{figure}
\begin{center}
\includegraphics{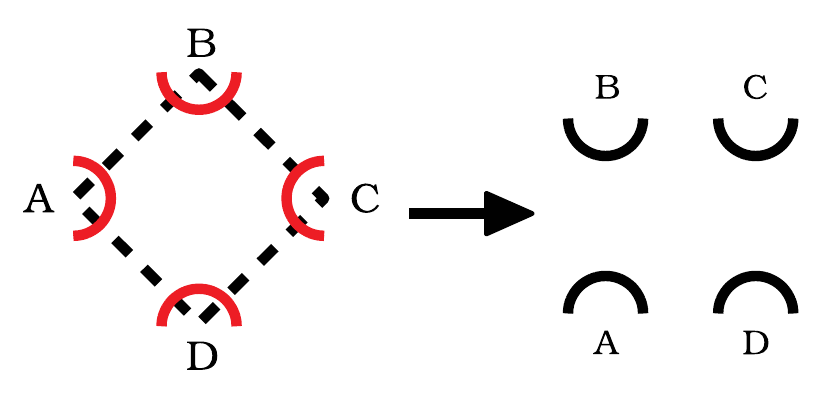}
\caption{The action of the operator $\mathrm{O}_0$. The incoming loop ends are at points A and D and the outgoing are at B and C. In words, the operation connects the two loop ends on either side of A and those on either side of D and then creates two small connected loop segments on either side of B and C. On the right is the schematic representation used to depict the other operators in figure \ref{fig:operators}. These representations correspond to the usual convention that the transfer ``time'' runs towards the North-East in the left part of the figure, due to the presence of the auxiliary space (and hence time runs upwards in the right part).}
\label{fig:O0}
\end{center}
\end{figure}

To complete a row, we first insert the extra loop ends corresponding to the auxiliary spaces. We then add the squares by operation with
\begin{equation}
\check{\sf R}_{2 n-2}...\check{\sf R}_4 \check{\sf R}_2 \check{\sf R}_0
\end{equation}
where each $\check{\sf R}_i$ adds a square at the position $i$. After all the squares are added, the auxiliary space is removed and the row is complete. An operator $\check{\sf R}_i$ is the sum of operators $\mathrm{O}_{\mathrm{P}}$ that correspond to each of the 14 planar partitions of the square illustrated in Figure \ref{fig:squares},
\begin{equation}
 \check{\sf R}_i=\sum_{\mathrm{P}} x_{\mathrm{P}} \mathrm{O}_{\mathrm{P}}
\end{equation}
where the sum is over all 14 planar partitions. The action of $\mathrm{O}_{\mathrm{P}}$ is on the four loop ends entering each grey square. In figure \ref{fig:O0}, we illustrate the action of the operator $\mathrm{O}_0$ (all four vertices disconnected) and give a schematic representation that we use in figure \ref{fig:operators} to show the action of all the operators. These can all be represented as products of TL operators\footnote{See \cite{Jacobsen14} and \cite{Jacobsen15} for further discussion. The relation to the TL operators is not particular important for us.}, $e_i$ ,  and these are also shown in figure \ref{fig:operators}.


\section{Parallel algorithm}
The transfer matrix approach is amenable to implementation in parallel. This was done in \cite{ScullardJacobsen2020} to reach $n=16$ for the Archimedean lattices and we will make some use of it here. During the computation, the data is a vector containing the weights of each connectivity state of loop ends that has appeared so far in the calculation. Careful consideration must be given to the way in which the states are distributed among processors. The ideal situation would be that upon operating with $\check{\sf R}_i$ on a state on processor $X$, a new state that appears as a result also resides on $X$ and no inter-processor communication would be needed. Unfortunately, it is impossible to decompose the problem in this way, and we will instead have to be content with an algorithm that approximates this ideal as closely as possible. For inspiration, we turn to Jensen's \cite{Jensen03} computation of the number of self-avoiding polygons (SAP) on the square lattice. The main idea of Jensen's algorithm is to divide the loop ends into two segments, the ones closest to the square $i$ on which one is currently operating and the other more distant ends. The goal is then to find some quantity that can easily be computed from the state of the distant loop ends that is unchanged upon an operation on the local loop ends, i.e., to find an invariant associated with the distant configuration. States with the same invariant are then placed on the same processor, and one is guaranteed not to have to communicate the result of the transfer operation. For the SAP problem, Jensen was able to find an exact invariant and to work perfectly in parallel. As far as we know, our problem does not have an exact invariant but we can find a partial invariant, and we nevertheless find it very useful to employ this general framework.

\begin{figure}
\begin{center}
 \includegraphics[width=12cm]{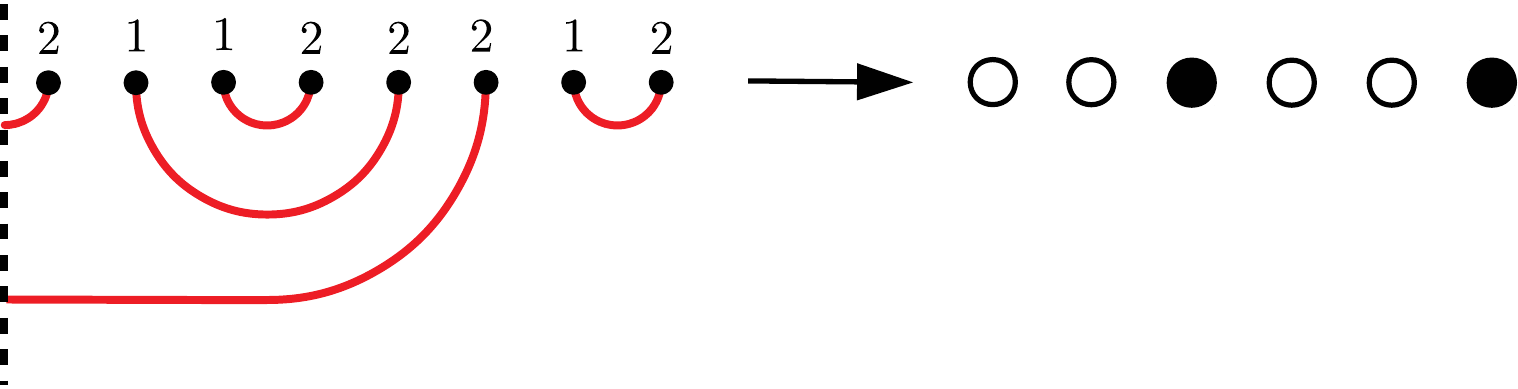}
 \caption{A loop configuration on a row segment and its corresponding black-white sites. Two immediate neighbours that are the endpoints of the same small arc are combined into a single black site. All the rest of the loop ends become white sites.}
 \label{fig:BW}
\end{center}
\end{figure}

\begin{figure}
\begin{center}
 \includegraphics[width=1.75in]{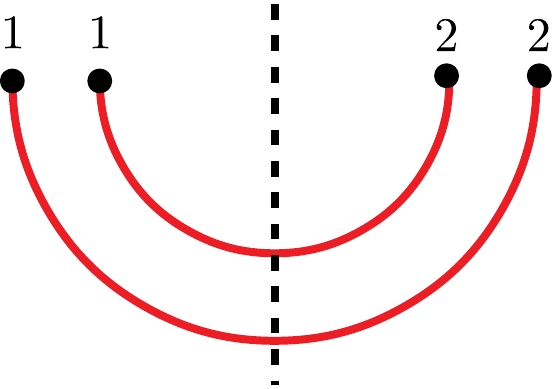}
 \caption{State of loop ends that can lead to the creation of a new black site. The ``distant'' loop ends are to the right and initially would contribute two white sites to the BW configuration. But joining the two loop ends on the left leads to a new black site on the right.}
 \label{fig:new_black}
\end{center}
\end{figure}

In our construction, we convert a loop configuration into a set of ``black'' and ``white'' sites in the following way. Two immediately-neighbouring loop ends for which the one on the left is 1 and the one on the right is 2, so they subtend the smallest possible arc, are combined to form a single black site. Each of the remaining loop ends is replaced with a white site, as shown in Figure \ref{fig:BW}. For $n$ loop ends, the number of sites in the BW configuration can vary from $n/2$ (all black) to $n$ (all white). And although every loop configuration corresponds to a black-white (BW) configuration, most BW configurations have many possible origins as loop states (the exception being all sites black). Now, let us divide our loop ends into two halves, left and right, and partially order the states according to their BW configuration in the right half. An application of $\check{\sf R}_i$ in the left half will generally not change the BW configuration of sites in the right. Unfortunately, it is not completely invariant because, while it is not possible to destroy an existing black site, it is possible for new black sites to appear under some circumstances. An example is shown in figure \ref{fig:new_black}, where joining the two loop ends on the left, or local, region turns two white sites into a single black site on the distant region. However, this is rare enough that placing states with the same right-half BW state on the same processor keeps the inter-processor communication to a minimum. In practice we divide the problem into more than two segments to allow us to efficiently use a large number of processors. The overview of the algorithm is as follows
\begin{enumerate}
 \item Insert the auxiliary spaces.
 \item Divide the sites into $s$ segments, and place all states with the same BW configuration {\it outside the first segment} onto the same processor.
 \item Carry out the transfer matrix operations, sending a few states between processors as needed, until reaching the segment boundary.
 \item Before crossing into the new segment, redistribute the vector such that all states with the same BW configuration outside of the segment about to be entered are on the same processor.
 \item Return to (ii) until reaching the end of the row.
 \item Remove the auxiliary spaces.
 \item Perform transfer operations without the auxiliary spaces, which may be needed for some lattices.
\end{enumerate}
This procedure is then iterated until the eigenvalue converges.

In our transfer matrix calculation, states are stored in a hash table that gains an entry when a new connectivity state is encountered; the final size of the vector is not exactly predictable, as the peculiarities of an individual lattice may not allow all possible connections. Thus, the calculation begins with only a single state, which then grows as the calculation proceeds, eventually leveling off when all realizable states have been encountered. When running in parallel, one would like not only to keep inter-processor communication to a minimum, but also to ensure that the load on each processor is in balance. So upon crossing into a new segment, we assign to each state a number $c$ computed from the BW configuration on the segments and take an inventory of the number, $N(c)$, of states with a given $c$. This must be done on each processor, with the results sent to processor 0 where the global sum is formed. The mapping between BW configurations and the integer $c$ is given in \ref{sec:integer_map}. To determine where to actually send each block of $N(c)$ states, we divide the calculation into three phases. 

In the first phase, the initial stage, there are many $c$ for which $N(c)=0$, and meticulously balancing the data at this point is not really worth the effort. So here we simply assign the data sequentially. That is, if we send the states $c$ to processor $j$, then we will send the states $c+1$ to $j+1$. 

In the second phase, which we define to start when $N(c) \neq 0$ for 80\% of the possible values of $c$, we use Jensen's load-balancing algorithm. This works as follows, which is slightly modified from Ref. \cite{Jensen03},
\begin{enumerate}
 \item Sort the global $N(c)$ on processor 0.
 \item On processor 0, assign each $c$ to the processors by first setting a processor identification counter $p_\mathrm{id}=0$ and then:
 \begin{enumerate}
 \item Among the unassigned $c$ values, assign the most frequent to $p_\mathrm{id}$. If $p_\mathrm{id}=0$, increase it by one. Otherwise, if the number of states on $p_\mathrm{id}$ is less than the number on processor 0, then also assign the next-most frequent $c$, and so on until the number of states on $p_\mathrm{id}$ exceeds those on 0.
 \item Increase $p_\mathrm{id}$, returning to $p_\mathrm{id}=0$ when we have passed the last processor.
 \item Return to (a) until all the states have been assigned.
 \end{enumerate}
 \item Send the list of owners of each $c$ to all the processors.
 \item On each processor, run through the hash table and send all states to their appropriate destinations.
\end{enumerate}
Our algorithm differs slightly from Jensen's in that in part (ii)(a) we always assign the most-frequent $c$, whereas Jensen first assigns the most-frequent and then uses the least-frequent configurations to fill in the gap with processor 0. We find that we achieve an acceptably balanced calculation with our variation.

In the third phase of the computation, the hash table has reached saturation and no new states are added upon performing transfer operations. This stage begins when we complete one full row without any new states (with the auxiliary space in) appearing. At this point, it would be wasteful to continue with the sorting and assignment algorithm outlined above because the result will no longer change, so we simply consult the ownership tables that were previously computed. Note that there are separate tables for each segment and for when the auxiliary space is in or out. The majority of the transfer matrix calculation occurs in this phase.

This was the parallel algorithm used in \cite{ScullardJacobsen2020} to compute thresholds up to $n=16$, utilizing millions of CPU hours. Here, we make only light use of parallel calculations; we use it for $n=10$ and $n=11$, which we do for only two points. To complete one Householder iteration takes about five hours on 72 processors for $n=11$ and we generally need about three iterations to get a converged value. As such, we used about 2000 to 2500 total CPU hours here. However, we have described the algorithm in detail here because it was not thoroughly discussed in \cite{ScullardJacobsen2020}.

\section{Results}
\subsection{Critical surface}
We have evaluated the curve $P_3(P_0)$ for values of $P_0$ with spacing $\Delta P_0=0.01$. The results are shown in table \ref{tab:P3}. These calculations were carried to at least $n=9$ in all cases, and for $P_0=0$ and $P_0=0.25$ we have used parallel computations to get to $n=11$. The raw data for every $n$ and $P_0$ is included in the Mathematica file {\verb Kagome_hypergraph_SJZ.nb } including as supplemental material to this paper. It is clear from the table, and from figure \ref{fig:P3} that the accuracy of the polynomial method depends on the proximity of the problem to the exact case, $P_0^*$, and we make this observation more quantitative below.

We have done these calculations to answer a variety of different questions about this system, and we begin with the finite-size scaling of the critical $P_3$ estimates.
\begin{table}
\begin{center}
 \begin{tabular}{l|l}
  $P_0$ & $P_3$ \\ \hline
  0    & $0.445\,905\,495\,45(3)$ \\
  0.01 & $0.454\,367\,078(1)$ \\
  0.02 & $0.462\,825\,465(2)$ \\
  0.03 & $0.471\,279\,300\,7(3)$\\
  0.04 & $0.479\,727\,306\,9(2)$\\
  0.05 & $0.488\,168\,287\,5(1)$\\
  0.06 & $0.496\,601\,120\,53(2)$\\
  0.07 & $0.505\,024\,756\,740(5)$\\
  0.08 & $0.513\,438\,216\,044(4)$\\
  0.09 & $0.521\,840\,584\,648\,5(3)$\\
  0.1  & $0.530\,231\,011\,951\,25(3)$\\
  0.11 & $0.538\,608\,707\,503\,6(4)$\\
  0.12 & $0.546\,972\,938\,019\,5(8)$\\
  0.13 & $0.555\,323\,024\,448(2)$\\
  0.14 & $0.563\,658\,339\,12(2)$\\
  0.15 & $0.571\,978\,303\,03(2)$\\
  0.16 & $0.580\,282\,383\,06(2)$\\
  0.17 & $0.588\,570\,089\,55(2)$\\
  0.18 & $0.596\,840\,973\,73(2)$\\
  0.19 & $0.605\,094\,625\,43(2)$\\
  0.2  & $0.613\,330\,670\,81(5)$\\
  0.21 & $0.621\,548\,770\,24(2)$\\
  0.22 & $0.629\,748\,616\,26(2)$\\
  0.23 & $0.637\,929\,931\,6(2)$\\
  0.24 & $0.646\,092\,467\,8(1)$\\
  0.25 & $0.654\,236\,002\,94(2)$ \\
  0.26 & $0.662\,360\,339\,8(2)$\\
  0.27 & $0.670\,465\,305\,7(2)$\\
  0.28 & $0.678\,550\,750\,0(5)$\\
  0.29 & $0.686\,616\,542\,8(2)$\\
  0.3 & $0.694\,662\,574\,2(2)$ \\
 \end{tabular}
 \caption{Extrapolations to infinite basis (i.e. $n \rightarrow \infty$) of the critical $P_3$ for the given values of $P_0$. All computations were taken to $n=9$, but with $P_0=0$ and $P_0=0.25$ taken to $n=11$.}
 \label{tab:P3}
\end{center}
\end{table}

\subsection{Scaling and extrapolation}
In previous work \cite{Jacobsen15,ScullardJacobsen2020}, it has been shown empirically that critical threshold estimates, $p_c(n)$, obtained by the critical polynomial method scale for large $n$ according to
\begin{equation}
 p_c(n) \approx p_c(\infty)+\sum_{k=1}^{\infty} \frac{a_k}{n^{\Delta_k}},
\end{equation}
and we order the $\Delta_k$ so that $\Delta_k > \Delta_{k'}$ for $k>k'$. Although there is no theoretical understanding of this formula, there is now ample numerical evidence that it is correct \cite{ScullardJacobsen2020,Jacobsen15}. In previous work, it was shown that for the kagome and $(3,12^2)$ lattices, both of which are cases of this hypergraph system, the lowest-order exponent is $\Delta_1=6$. It is natural, then, to hypothesize that this is in fact the case for the entire range of $P_0$ and we work to verify this here.

The version of this scaling relevant to the kagome hypergraph is
\begin{equation}
 P_3(P_0;n)=P_3(P_0;\infty)+f(P_0;n)+\sum_k^{\infty} \frac{a_k(P_0)}{n^{\Delta_k}} \label{eq:KHscaling}
\end{equation}
where we have explicitly inserted a correction term, $f(P_0;n)$, for small $n$ which must go to zero faster than any power of $n$ as $n \rightarrow \infty$. One thing to point out here is that at the exact point $P_0=P_0^*$, critical polynomials predict $P_3^*$ for all $n$ (the value is of course independent of $n$) and thus we must have
\begin{equation}
 f(P_0^*;n)=a_k(P_0^*)=0
\end{equation}
for all $n$ and $k$. 

\begin{figure}
\begin{center}
 \includegraphics[width=3.5in]{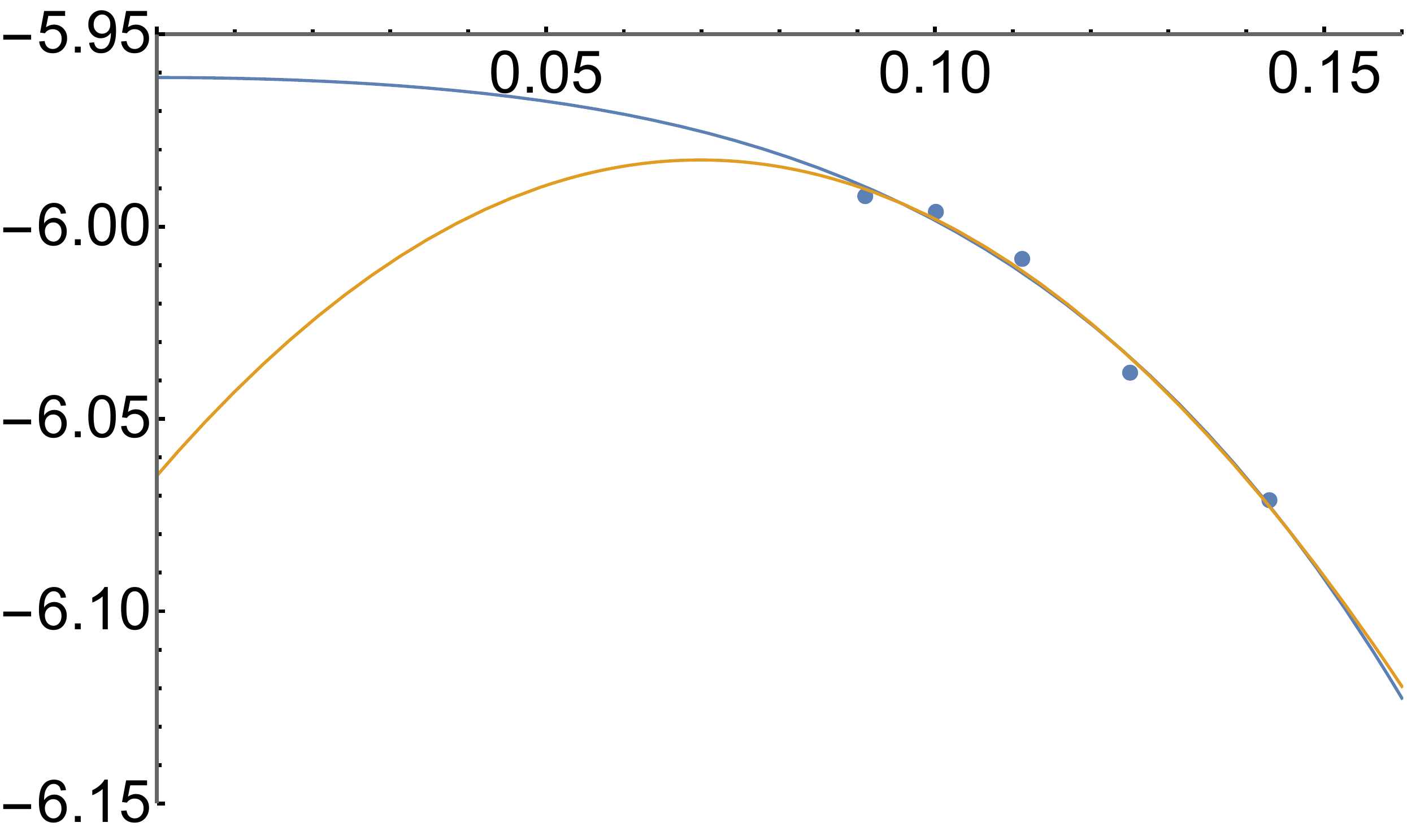}
 \caption{$\Delta_1(n)$ determined from (\ref{eq:qn}) for $P_0=0.25$ plotted against $1/n$. The blue (resp.\ orange) curve is a fit second-degree polynomial in the variable $1/n^2$ (resp.\ $1/n$) to the last five data points.}
 \label{fig:scaling}
\end{center}
\end{figure}

To determine $\Delta_1$, we follow the procedure outlined in \cite{ScullardJacobsen2020} and assume the truncated form
\begin{equation}
 P_3(P_0;n) \approx P_3(P_0;\infty)+\frac{a_1(P_0)}{n^{\Delta_1}} .
\end{equation}
We then form the quantities
\begin{equation}
 q(n) \equiv \frac{P_3(n)-P_3(n-1)}{P_3(n-1)-P_3(n-2)}
\end{equation}
which is a function only of $n$ and $\Delta_1$, i.e.,
\begin{equation}
 q(n) = \left(1-\frac{2}{n} \right)^{\Delta_1} \frac{n^{\Delta_1}-(n-1)^{\Delta_1}}{(n-1)^{\Delta_1}-(n-2)^{\Delta_1}} \label{eq:qn}
\end{equation}
which provides us with the estimates, $\Delta_1(n)$. Fitting these estimates to polynomials of various order in $1/n$, we can obtain a value for $\Delta_1$. In figure \ref{fig:scaling} we plot $\Delta_1(n)$ vs.\ $1/n$ for the case $P_0=0.25$. The blue (resp.\ orange) curve is a fit second-degree polynomial in the variable $1/n^2$ (resp.\ $1/n$) to the last five data points. Supposing those to delimit the range of reasonable extrapolations (a hypothesis that seems to agree with the visual inspection of the figure), we arrive at the final value $\Delta_1 = 6.01 (5)$. This was the same value found for the kagome and $(3,12^2)$ lattices \cite{ScullardJacobsen2020}. We therefore conjecture that $\Delta_1$ is constant over the entire range of $P_0$ and that the same is true for the other exponents, $\Delta_k$, which thus take the values previously determined \cite{Jacobsen15,ScullardJacobsen2020} for the kagome and $(3,12^2)$ lattices, $\Delta_2=7$, $\Delta_3=8$, etc.

It is an open challenge to gain a theoretical understanding of the scaling (\ref{eq:KHscaling}). In work by one of us with Mertens \cite{Mertens2016}, the critical polynomial was given an alternative definition in which it was related to the cluster density. The scaling theory of that quantity has been studied more extensively \cite{Mertens2017,ZiffFinchAdamchik} than that of critical polynomials and perhaps this will eventually shed light on the problem. Or perhaps tools from conformal field theory will prove to be decisive here. But whatever the eventual path looks like, it is safe to say that at present the problem has barely been touched.

\subsection{Derivatives at $P_0^*$} \label{sec:derivatives}
If we could accurately evaluate the derivatives of $P_3(P_0)$ at $P_0=P_0^*$, we could use this knowledge to construct an approximation to the entire curve by using its power series,
\begin{equation}
P_3(P_0)=P_3^*+P_3'(P_0^*)(P_0-P_0^*)+\frac{1}{2}P_3''(P_0^*)(P_0-P_0^*)^2 \cdots \ .
\end{equation}
Because the critical polynomial estimates become more accurate the closer we choose $P_0$ to $P_0^*$, i.e., the smaller we make $\epsilon \equiv P_0-P_0^*$, one might think that we can estimate derivatives with arbitrary precision. Unfortunately, this proves not to be the case in general. However, starting with the first derivative and using the approximate central difference formula
\begin{equation}
 P_3'(P_0^*) \approx \frac{P_3(P_0^*+\epsilon)-P_3(P_0^*-\epsilon)}{2 \epsilon}+\mathrm{O}(\epsilon^2) \label{eq:CD1}
\end{equation}
we find the estimates in Table \ref{tab:firstderiv} for $\epsilon=10^{-20}$. We appear to have 40 converged digits already at $n=1$. It seems clear then that the $n=1$ result is accurate to order $\epsilon^2$. This makes computing the first derivative basically trivial; all we need is $n=1$, meaning that this derivative can be found exactly by differentiating (\ref{eq:firstpoly}). In Ref. \cite{ZiffGu09} it was postulated by a different argument that this derivative is given exactly by
\begin{equation}
 P_3'(P_0^*)=\frac{1}{2-p_c^{\mathrm{DBH}}} \label{eq:P3deriv}
\end{equation}
and this is consistent with (\ref{eq:firstpoly}) and the numerical results in Table \ref{tab:firstderiv}. Note that the subsequent $n$ in Table \ref{tab:firstderiv} do not add any digits of accuracy because the finite difference formula (\ref{eq:CD1}) is itself only good to $\mathrm{O}(\epsilon^2)$. We could rectify this by using a higher-order formula, but if we really wanted more digits we could just make $\epsilon$ smaller.

Moving on to the second derivative, the picture is different. In Table (\ref{tab:secondderiv}), we show the result of using the second-order central difference formula
\begin{equation}
 P_3''(P_0^*) \approx \frac{P_3(P_0^*-\epsilon)-2 P_3(P_0^*)+P_3(P_0^*-\epsilon)}{\epsilon^2} \label{eq:CD2}
\end{equation}
and it seems that by $n=7$ we have perhaps four converged digits, with the $n=1$ estimate nowhere near the correct value. It seems that computing the second derivative is more difficult due to the scaling of the accuracy of the points $P_3(P_0^*-\epsilon)$ with $\epsilon$, and cannot be found exactly as can the first. The same is true of the higher derivatives and in fact the problem of slow convergence with $n$ gets progressively worse with increasing order.

Although disappointing, this result was perhaps to be expected. If all the derivatives were as easy to compute as the first, this would constitute an exact solution for the critical surface, which would in one stroke solve many unsolved problems, such as bond percolation on the kagome lattice and site percolation on the hexagonal lattice. The apparently fundamental difficulty of computing these higher derivatives would seem to be evidence that these problems are not solvable. However, we will see later that fitting the coefficients of a Taylor expansion of $P_3(P_0)$ about $P_0^*$ gives us a very accurate approximation, so we will not press the direct computation of derivatives any further. Note that the second derivative was also estimated in \cite{ZiffGu09} as -0.11974, which was based on a fit to only a handful of points and is thus reasonably close to our more precise estimate. 

\begin{table}
\begin{center}
 \begin{tabular}{l|l}
  $n$ & $P_3'(P_0^*)$ \\ \hline
  1 & 0.838\,856\,338\,383\,645\,498\,540\,759\,130\,490\,605\,223\,447\,275\,665\,164\,73 \\
  2 & 0.838\,856\,338\,383\,645\,498\,540\,759\,130\,490\,605\,223\,447\,225\,361\,665\,82 \\
  3 & 0.838\,856\,338\,383\,645\,498\,540\,759\,130\,490\,605\,223\,447\,241\,868\,263\,57 \\
  4 & 0.838\,856\,338\,383\,645\,498\,540\,759\,130\,490\,605\,223\,447\,244\,407\,526\,56 \\
  5 & 0.838\,856\,338\,383\,645\,498\,540\,759\,130\,490\,605\,223\,447\,244\,879\,199\,85 \\
  6 & 0.838\,856\,338\,383\,645\,498\,540\,759\,130\,490\,605\,223\,447\,244\,992\,855\,73 \\
  7 & 0.838\,856\,338\,383\,645\,498\,540\,759\,130\,490\,605\,223\,447\,245\,026\,507\,61 \\
 \end{tabular}
 \caption{Estimates of the derivative $P_3'(P_0^*)$ for bases of width $n$ using the central difference formula (\ref{eq:CD1}) with $\epsilon=10^{-20}$.}
 \label{tab:firstderiv}
\end{center}
\end{table}

\begin{table}
\begin{center}
 \begin{tabular}{l|l}
  $n$ & $P_3''(P_0^*)$ \\ \hline
  1 & 0.707\,609\,089\,804\,785\,291\,906\,011\,135\,318 \\
  2 & -0.084\,596\,936\,608\,756\,411\,804\,148\,753\,338 \\
  3 & -0.120\,887\,799\,463\,767\,062\,854\,390\,279\,638 \\
  4 & -0.124\,063\,709\,127\,716\,798\,971\,148\,182\,541 \\
  5 & -0.124\,537\,426\,847\,330\,612\,081\,080\,233\,991 \\
  6 & -0.124\,641\,324\,029\,672\,773\,417\,692\,626\,285 \\
  7 & -0.124\,670\,864\,259\,082\,801\,908\,587\,827\,416 \\
 \end{tabular}
 \caption{Estimates of the second derivative $P_3''(P_0^*)$ for bases of width $n$ using the central difference formula (\ref{eq:CD1}) with $\epsilon=10^{-20}$.}
 \label{tab:secondderiv}
\end{center}
\end{table}

\subsection{Accuracy of critical polynomials}
As shown in the previous section, if we seek an estimate for a system a distance $\epsilon \equiv |P_0-P_0^*|$ from the exact solution, then the $n=1$ polynomial estimate is accurate to $\epsilon^{-2}$ digits. There is currently no theoretical understanding of this, but in terms of the scaling (\ref{eq:KHscaling}) we must have
\begin{eqnarray}
 f(P_0;L) &\sim& \epsilon^{m_0} \\
 a_k(P_0) &\sim& \epsilon^{m_k}
\end{eqnarray}
as $\epsilon \rightarrow 0$, where $m_k \ge 2$ for all $k$.

If we now consider the kagome lattice, using our estimate for the critical probability, $p_c$, we find $P_0(\mathrm{kagome})=0.107\,575\,125\,970\,732$
and $\epsilon \approx 0.011\,047$. Because $\epsilon^2 \sim 10^{-4}$, we expect the $n=1$ polynomial estimate to be accurate to around four digits, which seems to be the case. Turning to the $(3,12^2)$ lattice, we find $\epsilon \approx 0.003\,927$ and $\epsilon^2 \sim 10^{-5}$ so we expect to start with five digits and indeed, this is once again the case. This is why we were able to obtain an extremely accurate estimate for this lattice by going to $n=16$ \cite{ScullardJacobsen2020}.

\begin{figure}
\begin{center}
\includegraphics[width=4.5in]{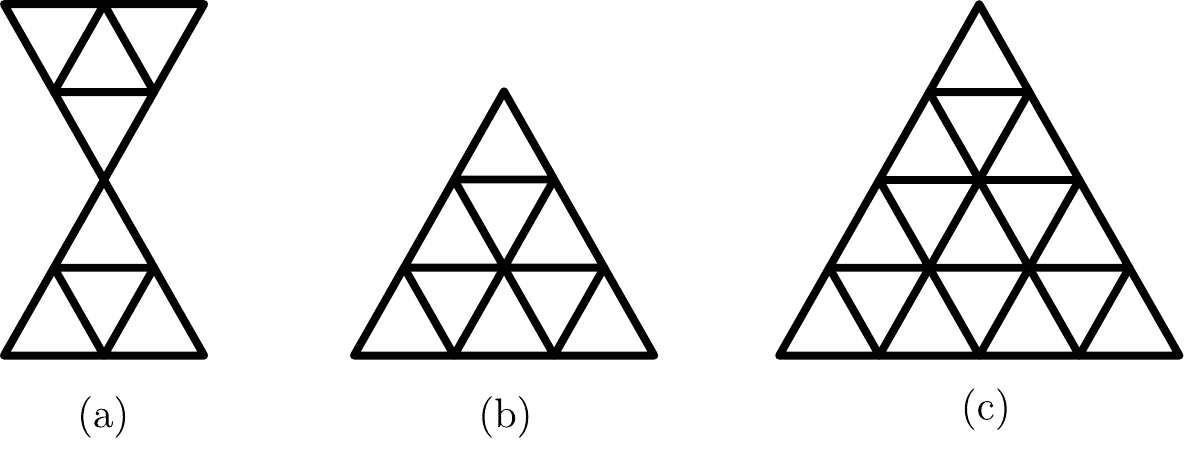}
\caption{(a) unit cell for the kagome subnet 2; (b) the triangle for subnet 3; (c) the triangle for subnet 4.}
\label{fig:subnets}
\end{center}
\end{figure}

In order to calculate $\epsilon$ for a given lattice, we must first know $P_0$ and thus $p_c$, and therefore one does not know in advance how accurate to expect the critical polynomial estimate to be. On the other hand, a fact pointed out in \cite{ZiffGu09} provides us with a simple rule. Consider the kagome hypergraph with the triangles shown in figure \ref{fig:subnets}, termed ``subnets'' in Refs.\ \cite{ZiffGu09} and \cite{Wu10}; subnet 3 is shown in figure \ref{fig:subnets}a and subnet 4 is in figure \ref{fig:subnets}(c). For subnet $m$ with $m$ very large, the interior of a triangle, far from the boundaries, is an ordinary triangular lattice. The threshold of the whole system is far greater than $p_c(\mathrm{tri})=2 \sin \pi/18$, so at criticality each triangle is supercritical, with a very large cluster existing in its center. Each corner can therefore connect with the others only through touching this central cluster, and we denote the probability of this $P_{\infty,\mathrm{corner}}$ in the limit $m \rightarrow \infty$. The system is thus effectively the double-bond hexagonal lattice, with the critical point given by
\begin{equation}
 P_{\infty,\mathrm{corner}}=p_c^{\mathrm{DBH}}
\end{equation}
and $P_0$ and $P_3$ become $P_0^*$ and $P_3^*$. 

This same logic holds whether the interior of a triangle contains a simple triangular lattice or something more complicated. This leads to a fairly simple rule: the more edges we pack into a triangle of the kagome hypergraph, the more accurate the first critical polynomial estimate will be. This explains why the bond percolation threshold of the $(3,12^2)$ lattice was found with such accuracy by this method; it has 4.5 edges per triangle whereas for the kagome lattice we have only 3. We should expect, then, that the $n=1$ polynomial estimate for the subnet 4 of figure \ref{fig:subnets}c will be very accurate indeed, as each triangle contains 30 edges. Interestingly, this calculation has been done already. Wu's ``homogeneity'' approximation for kagome hypergraphs, which had its origins in a 1979 paper \cite{Wu79} and was explored by him and collaborators more recently \cite{Wu10,Ding10}, turns out to give estimates identical to those found with the $n=1$ critical polynomials. Because there are 30 edges in a triangle, the calculation is rather involved, but it was carried through in \cite{Wu10}, with the estimate $p_c=0.625\,364\,661\,497\,144$. The earlier Monte Carlo estimate for this system performed in \cite{ZiffGu09}, $p_c = 0.625365(3)$, was not sufficient to rule out this number. The more precise $p_c=0.625\,364\,24(7)$ \cite{Ding10} was needed to show that the $n=1$ estimate fails in the seventh digit. Note that although packing the triangle with bonds and sites will push the critical system towards the exact point, $P_0^*$, this does not necessarily allow us to calculate the limiting $p_c$ analytically. The problem is that although the $n=1$ polynomial goes to the exact answer in the limit $m \rightarrow \infty$, with many bonds and/or sites in one triangle it becomes increasingly difficult to calculate $P_3$ and $P_0$ as functions of $p$. Indeed, for the subnet 4 calculations done in \cite{Wu10}, calculating (the equivalent of) these quantities on a single triangle had to be done by a computer algorithm. As $m$ becomes large, the computational complexity is shifted to this problem rather than that of computing the eigenvalues.

\subsection{Critical surface fit}
Armed with the above results, we are now in a position to provide a fit for the function $P_3(P_0)$. We will expand around the exact point, $P_0^*$, as suggested above, i.e.,
\begin{equation}
 P_3(P_0) \approx P_3^*+\sum_{k=1}^N \frac{a_k}{k!}(P_0-P_0^*)^k \label{eq:P3fit}
\end{equation}
and we will fit the $a_k$ to the data in table \ref{tab:P3}, rather than trying to compute them as derivatives. Because $a_1=P_3'(P_0^*)$ is known exactly (see eq.\ (\ref{eq:P3deriv}) ), we need to fit $N-1$ parameters. If $N$ is chosen too small, the formula may be inaccurate away from $P_0^*$. For example, if we choose $N=1$, keeping only the exact terms, for the kagome lattice we find $p_c \approx 0.524\,408\,77$ \cite{ZiffGu09}, a reasonable approximation that was once conjectured to be the exact solution \cite{Sakamoto}. However, this level of approximation gives for the site hexagonal problem $p_c \approx 0.698\,914\,02$, a comparatively poor estimate.
On the other hand, with $N$ too large we may have too many fitting parameters. We optimize this by finding the $N$ such that the predicted critical $P_3$ furthest from $P_0^*$, i.e., for the site threshold of the hexagonal lattice, is unchanged from $N-1$. To do this, we use Mathematica's {\verb FindFit[] } function for various $N$ and solve the equation
\begin{equation}
 p=P_3(1-p)
\end{equation}
for $p$ using the fitted $P_3(P_0)$. We find that fits done with $N=8$ and $N=9$ are the first pair to make identical predictions. The prediction is in fact quite stable for larger $N$ as well but with $N=11$ there is an additional small solution for real $p \in [0,1]$, indicating that we have too many terms. Additionally, the predictions for $a_2$ between $N=8$ and $N=9$ agree to 8 digits whereas the agreement between $N=9$ and $N=10$ is slightly worse. We thus settle on $N=9$ and report the fit coefficients in table \ref{tab:firstderiv}. Note that the fit for $a_2$ is in agreement with our direct computation shown in table \ref{tab:secondderiv}. Our prediction for the site hexagonal threshold is
\begin{equation}
 p_c(\mathrm{site\ hex})=0.697\,040\,220(5)
\end{equation}
where the error bar reflects the fact that two of our extrapolated points in table \ref{tab:P3} are only good to 9 digits, as well as the variation in this value seen with smaller $N$. Note that our value cannot be ruled out by the traditional numerical result given in (\ref{eq:hex_site_num}) \cite{FengDengBlote08}. If we use our formula to find the kagome bond threshold, the value of which we did not actually include in the fit, by solving the equation
\begin{equation}
 p^3+3p^2(1-p)=P_3([1-p]^3)
\end{equation}
we find $p_c=0.524\,404\,999\,2$, which agrees with the numerical value (\ref{eq:kagome_num}) to all 10 digits.

It is rather remarkable that with the eight coefficients in table \ref{tab:fit_coeffs} and the relatively simple formula (\ref{eq:P3fit}), we can find the threshold of any system along the curve $P_3(P_0)$ to a precision that rivals Monte Carlo and methods of similar accuracy. Another example we can study is mixed site-bond percolation on the hexagonal lattice. Here, sites are occupied with probability $p_s$ and bonds open with probability $p_b$ and if a site is unoccupied then any path of open bonds going through it becomes disconnected. The point $p_s=1$ corresponds to ordinary bond percolation and $p_b=1$ is pure site percolation. The triangular cell for this problem is like that shown in figure \ref{fig:3-12}(b) but where the interior triangle is contracted to a single site that has probability $p_s$ and the bonds have probability $p_b^{1/2}$. The functions $P_0(p_b,p_s)$ and $P_3(p_b,p_s)$ are given by \cite{ZiffGu09}
\begin{eqnarray}
 P_0(p_b,p_s)&=&1-p_s+p_s \left[\left(1-\sqrt{p_b} \right)^3+3\left(1-\sqrt{p_b} \right)^2 \sqrt{p_b} \right] \\
 P_3(p_b,p_s)&=&p_s p_b^{3/2}.
\end{eqnarray}
In table \ref{tab:p_b} (table \ref{tab:p_s}) we give some examples of critical $p_b$ ($p_s$) for given $p_s$ ($p_b$) calculated with our fit compared to the Monte Carlo results of \cite{Tarasevich1999}. Once again we find that our fit performs better than the Monte Carlo, although in this case the test is not particularly rigorous. For the homogeneous system $p_b=p_s$, our fit gives 
\begin{equation}
p_c=0.821\,722\,96(1)\ .
\end{equation}
The approximate formula in \cite{ZiffGu09} gives $p_c \approx 0.82199$ but we know of no other numerical results for this quantity.

\begin{table}
\begin{center}
\sisetup{table-figures-integer=3,table-figures-decimal=9,table-figures-exponent=4}
 \begin{tabular}{c|S[table-number-alignment=left]}
  $k$ & $a_k$ \\
  \hline
  2 & -0.1246880632 \\
  3 & -0.7811542720 \\
  4 & 5.817225212 \\
  5 & -29.75063849 \\
  6 & 60.23553836 \\
  7 & 1.166617815e3 \\
  8 & -1.265967276e4 \\
  9 & -2.557340356e3
 \end{tabular}
 \caption{Fit of the data in table \ref{tab:P3} to the form in equation (\ref{eq:P3fit}).}
 \label{tab:fit_coeffs}
 \end{center}
\end{table}

\begin{table}
\begin{center}
 \begin{tabular}{c|c|c}
  $p_s$ & $p_b$ & Ref.\ \cite{Tarasevich1999} \\ \hline
  0.80 & 0.848\,242\,90(5) & 0.8481(5) \\
  0.85 & 0.789\,506\,35(5) & 0.7890(5)  \\
  0.90 & 0.738\,142\,50(5) & 0.7377(5) \\
  0.95 & 0.692\,875\,82(5) & 0.6926(5)  \\
 \end{tabular}
 \caption{Critical values of $p_b$ for given $p_s$ for site-bond percolation on the hexagonal lattice, computed with the fit to equation (\ref{eq:P3fit}) and compared with Monte Carlo results \cite{Tarasevich1999}.}
 \label{tab:p_s}
 \end{center}
\end{table}
\begin{table}
\begin{center}
 \begin{tabular}{c|c|c}
  $p_b$ & $p_s$ & Ref.\ \cite{Tarasevich1999} \\ \hline
  0.80 & 0.840\,548\,52(5) & 0.8401(5) \\
  0.85 & 0.798\,606\,79(5) & 0.7986(5) \\
  0.90 & 0.761\,153\,88(5) & 0.7610(5) \\
  0.95 & 0.727\,486\,12(5) & 0.7275(5) \\
 \end{tabular}
 \caption{Critical values of $p_s$ for given $p_b$ for site-bond percolation on the hexagonal lattice, computed with the fit to equation (\ref{eq:P3fit}) and compared with Monte Carlo results \cite{Tarasevich1999}.}
 \label{tab:p_b}
\end{center}
\end{table}

\section{Conclusion}
We have studied percolation on the kagome hypergraph system using an adaptation of the method of critical polynomials. By computing critical values of $P_3$ for over 30 points along the critical surface, we were able to produce an eight-parameter fit capable of giving the critical point of any system lying on that surface to accuracy that rivals the best Monte Carlo. We also found that the accuracy of the critical polynomial method on the kagome hypergraph depends on the proximity of the system under consideration to the exact point, $P_0^*$. This explains why the bond percolation thresholds of certain lattices, such as $(3,12^2)$, could be computed to such high precision.

\section{Acknowledgements}
The work of CRS was performed under the auspices of the U.S. Department of Energy at the Lawrence Livermore National Laboratory under Contract No.\ DE-AC52-07NA27344 and was supported by the LLNL-LDRD Program under Project No.\ 19-DR-013. The work of JLJ was supported in part by the ERC advanced grant ``NuCFT''.

\section*{References}
\bibliographystyle{iopart-num}
\bibliography{SJZ}

\appendix

\section{Weights for transfer matrix computation} \label{sec:weights}
\begin{eqnarray}
x_0 &=& P_0^2+4 P_0 P_2 \\
x_4 &=& P_3^2 \\
x_{ABC} &=& P_3 P_2 \\
x_{BCD} &=& P_3 P_2 \\
x_{ADC} &=& P_3 P_2 \\
x_{ADB} &=& P_3 P_2 \\
x_{AB} &=& P_3 P_0+P_2 (P_0+2 P_2) \\
x_{BC} &=& P_2^2 \\
x_{CD} &=& P_3 P_0+P_2 (P_0+2 P_2) \\
x_{AD} &=& P_2^2 \\
x_{AB|CD} &=& P_2^2+2 P_3 P_2 \\
x_{AD|BC} &=& 0 \\
x_{BD} &=& P_2^2 \\
x_{AC} &=& P_2^2
\end{eqnarray}
\begin{figure}
\begin{center}
\includegraphics[width=5in]{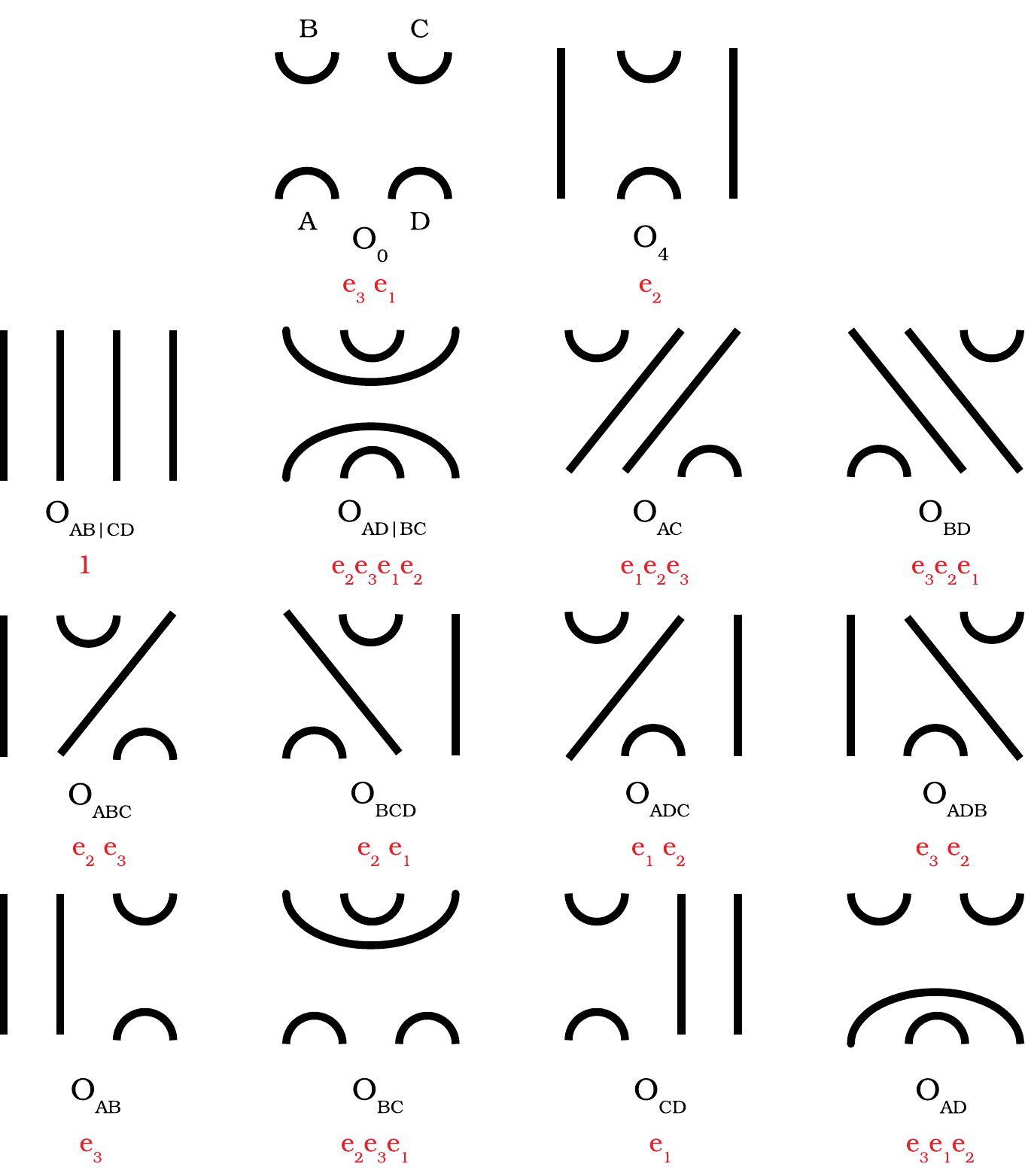}
\caption{The effect of the 14 planar connectivity operators on the loop ends connecting a square. The corresponding products of Temperley-Lieb operators, $e_i$, are shown below in red. For the purposes of the TL operators, the vertex A has loop ends labelled 1 and 2 and D has 3 and 4.}
\label{fig:operators}
\end{center}
\end{figure}

\section{Numbering black-white configurations} \label{sec:integer_map}
Here, we describe the mapping between black-white configurations and the integer $c$. As discussed in the main text, a pair of loop ends consisting of a $1$ with its partner $2$ on its immediate right are referred to as ``black'' ends. All other ends are called ``white''. We combine the two black loop ends into a single black site, so that if we have a system of $N_T$ loop ends and $n_b$ black sites then we have $N_T-2 n_b$ white sites and $N_B \equiv N_T-n_b$ total sites. Therefore, the number of possible configurations for a fixed $n_b$ is given by
\begin{equation}
 \Omega(n_b,N_T)= {N_T-n_b \choose n_b}
\end{equation}
The maximum number of black sites we can have is clearly
\begin{equation}
 n^{\mathrm{max}}_b= \left \lfloor \frac{N_T}{2} \right \rfloor
\end{equation}
and so for a given $N_T$, the total number of black-white states is
\begin{equation}
 N_{\mathrm{bw}}=\sum_{n_b=0}^{\left \lfloor N_T/2 \right \rfloor}\Omega(n_b,N_T)
\end{equation}
In order to map the states to integers, we will first develop such a map for a fixed number of black sites. For a fixed $n_b$, we do this in lexicographical order. Specifically, we will take the state with all black sites in their leftmost position to be state $0$. Then we add one to the state whenever we move the rightmost black site one position to the right. When this state can no longer be moved, we move the rightmost black site that can be moved one position, increment the state by one and place all black sites that are to right of the one just moved directly on its right. The final state is reached when all black sites are on the right. If $\{\sigma\}$ is the configuration of black and white sites, we define $S(\{\sigma\},n_b)$ to be the state number of a configuration with $n_b$ black sites fixed. As mentioned, $S(\{\sigma\},n_b,N_B)=0$ when all the black states are on the left. Now, we use $l$ to denote the position of the left-most black site in the system. Then clearly $S(\{\sigma\},1)=l$, but the real question is, when $n_b>1$, what is the minimum number of states that have been counted? We denote this quantity $M(N_B,n_b,l)$. Now, $M(N_B,n_b,0)=0$ because we cannot be sure any of the black sites have moved yet. If $l=1$, then we know that the $n_b-1$ black sites to the right of the first one have travelled through all their configurations on the $N_B-1$ bits, and so
\begin{equation}
 M(N_B,n_b,1)=\Omega(n_b-1,N_B-1) .
\end{equation}
It is therefore easy to see that
\begin{equation}
 M(N_B,n_b,l)=\sum_{i=1}^l \Omega(n_b-1,N_B-i) .
\end{equation}
To complete the assignment of an integer to a given configuration we can proceed recursively. If $S(\{\sigma\},n_b,N_B)$ is the state number then
\begin{equation}
 S(\{\sigma\},n_b,N_B)=M(N_B,n_b,l)+S(\{\sigma\}\setminus \sigma_0,n_b-1,N_B-l-1)
\end{equation}
To be completely explicit, if we have $N_T$ loop ends then we really want
\begin{equation}
 S(\{\sigma\},n_b,N_T-n_b)=M(N_T-n_b,n_b,l)+S(\{\sigma\}\setminus \sigma_0,n_b-1,N_T-n_b-l-1) .
\end{equation}
Now we have the state number for a given $n_b$. However, to define the total state number, we should first add the total number of states with black sites less than $n_b$, and thus
\begin{equation}
 S_T(\{\sigma\},N_T)=\sum_{i=0}^{n_b-1} {N_T-i \choose i}+S(\{\sigma\},n_b,N_T-n_b) .
\end{equation}

\end{document}